\documentclass[sigconf,authorversion,nonacm]{acmart}

\usepackage{amsmath}
\usepackage{booktabs} 
\usepackage{verbatim}
\usepackage{graphics}
\usepackage{graphicx}
\usepackage{epsfig} 
\usepackage{epstopdf}
\usepackage{subfigure}
\usepackage{balance}
\usepackage{comment}
\usepackage{booktabs}

\usepackage{amssymb}
\usepackage{amsfonts}
\usepackage{bm}
\usepackage{colortbl}
\usepackage{tabularx}
\usepackage{color}
\usepackage{xspace}
\usepackage{url}         
\usepackage{algorithm}
\usepackage{algpseudocode}
\usepackage{placeins}
\usepackage{multirow}
\usepackage{enumitem}
\usepackage{leading}
\usepackage{pgfplots}
\usepackage{xcolor}
\usepackage{soul}
\usepackage[amssymb]{SIunits}
\setcitestyle{numbers,sort&compress}

\DeclareMathOperator*{\argmin}{arg\,min}

\newcommand{\SystemName}{InaudibleKey\xspace}
\AtBeginDocument{%
  \providecommand\BibTeX{{%
    \normalfont B\kern-0.5em{\scshape i\kern-0.25em b}\kern-0.8em\TeX}}}

\setcopyright{acmcopyright}
\copyrightyear{2020}
\acmYear{2020}
\acmDOI{10.1145/1122445.1122456}

\acmConference[IPSN '21]{IPSN '21: Proceedings of the International Conference on Information Processing in Sensor Networks}{May 18-21, 2021}{Tennessee, USA}
\acmBooktitle{IPSN '21: Proceedings of the International Conference on Information Processing in Sensor Networks,
  May 18-21, 2021, Tennessee, USA}
\acmPrice{15.00}
\acmISBN{978-1-4503-XXXX-X/18/06}

\begin{document}

\title[]{\SystemName : Generic Inaudible Acoustic Signal based Key Agreement Protocol for Mobile Devices}


\author{Weitao Xu}
\email{weitaoxu@cityu.edu.hk}
\affiliation{%
  \institution{City University of Hong Kong}
	\country{Hong Kong SAR, China}
}

\author{Zhenjiang Li}
\affiliation{%
  \institution{City University of Hong Kong}
	\department{and City University of Hong Kong}
	\country{Shenzhen Research Institute, China}
}

\author{Wanli Xue$^\dagger$}
\email{wanli.xue@cybersecuritycrc.org.au}
\affiliation{%
  \institution{University of New South Wales}
  \additionalaffiliation{Cyber Security Cooperative Research Centre (CSCRC), Australia}
	\country{Australia}
}

\author{Xiaotong Yu}
\email{xiaotong.yu96@gmail.com}
\affiliation{%
  \institution{University of New South Wales}
	\country{Australia}
}

\author{Bo Wei}
\email{bo.wei@northumbria.ac.uk}
\affiliation{%
  \institution{Northumbria University}
	\country{UK}
}

\author{Jia Wang}
\email{jia.wang@szu.edu.cn}
\affiliation{%
  \institution{Shenzhen University}
	\country{China}
}

\author{Chengwen Luo}
\email{chengwen@szu.edu.cn}
\affiliation{%
  \institution{Shenzhen University}
	\country{China}
}

\author{Wei Li}
\email{weiwilson.li@sydney.edu.au}
 \affiliation{
  \institution{The University of Sydney}
	\country{Australia}
}

\author{Albert Y. Zomaya}
\email{albert.zomaya@sydney.edu.au}
 \affiliation{
  \institution{The University of Sydney}
	\country{Australia}
}

\renewcommand{\shortauthors}{W. Xu et al.}
\begin{abstract}
Secure Device-to-Device (D2D) communication is becoming increasingly important with the ever-growing number of Internet-of-Things (IoT) devices in our daily life. To achieve secure D2D communication, the key agreement between different IoT devices without any prior knowledge is becoming desirable. Although various approaches have been proposed in the literature, they suffer from a number of limitations, such as low key generation rate and short pairing distance. In this paper, we present InaudibleKey, an inaudible acoustic signal based key generation protocol for mobile devices. Based on acoustic channel reciprocity, InaudibleKey exploits the acoustic channel frequency response of two legitimate devices as a common secret to generating keys. InaudibleKey employs several novel technologies to significantly improve its performance. We conduct extensive experiments to evaluate the proposed system in different real environments. Compared to state-of-the-art works, InaudibleKey improves key generation rate by 3-145 times, extends pairing distance by 3.2-44 times, and reduces information reconciliation counts by 2.5-16 times. Security analysis demonstrates that InaudibleKey is resilient to a number of malicious attacks.  We also implement InaudibleKey on modern smartphones and resource-limited IoT devices. Results show that it is energy-efficient and can run on both powerful and resource-limited IoT devices without incurring excessive resource consumption.
\end{abstract}




\maketitle
\section{Introduction}
\label{sec:introduction}
\subsection{Background}
With recent advances in mobile computing and embedded technology, there is an increasing number of IoT devices in our daily life, such as smartphone, smart watch, and Google assistant. Correspondingly, it is more and more common to pair two devices for the purpose of data sharing, synchronization, and collaboration. For example, two persons that meet for the first time in a meeting want to associate their smartphones temporarily to exchange their name card. Due to the "open air" nature of wireless communication, cryptographic key agreement is a fundamental requirement to secure D2D communication to achieve confidentiality~\cite{xi2014keep,xu2021key}.

The secure key distribution between two communication parties can be addressed by public key infrastructure (PKI). Unfortunately, public key-based solutions are not applicable to mobile devices because PKI only works if the identity of the other party is known out of band or only trusted parties have identities signed by pre-established certificate authorities. Another solution is pre-distributed key which is usually in the form of master key or key material. However, key pre-distribution schemes lack scalability, which makes them incapable in a dynamic environment, where new devices may join and quit frequently. Near field communication (NFC) is becoming more and more popular in modern mobile devices, but its communication range is limited to only tens of centimetres (typically <20~cm). Diffie-Hellman protocol (also named D-H protocol) is a popular key establishment protocol to establish cryptographic keys over a public channel. However, D-H protocol is susceptible to man-in-the-middle (MITM) attack and authenticated D-H protocol requires the presence of certificate authority (CA). 
The most common method to pair mobile devices is still asking the user to scan nearby devices, choose the target device and confirm manually, which is neither user friendly nor suitable for devices without screens.
\vspace{-0.15in}
\subsection{Motivation}
The lack of efficient and user-friendly pairing methods has inspired researchers to explore suitable alternatives to authenticate mobile devices. Because two devices are not assumed to own any shared knowledge as a-priori, the widely adopted design principle in the literature is that if multiple devices share a similar observation of certain random signal, then the signal can be used to extract keys. Different designs explore different forms of such random signals~\cite{jana2009effectiveness,xi2016instant,mathur2008radio,liu2013fast,varshavsky2007amigo,mathur2011proximate,karapanos2015sound,xu2016walkie,shen2018shake}, while they all strive to achieve a \textit{fast} (sufficient bit generation rate), \textit{practical} (without requiring extra hardware) and \textit{ubiquitous} (usable in different environments) key generation system design.

A successful set of pioneer efforts is to leverage wireless channel information~\cite{mathur2008radio,jana2009effectiveness,xi2014keep,liu2013fast,xi2016instant,wang2011fast}, such as Received Signal Strength Indicator (RSSI) and Channel State Information (CSI). These methods are based on wireless channel reciprocity which means the channel characteristics (RSSI or CSI) measured between two devices by exchanging a pair of probe packets in quick succession, will be nearly the same. However, RSSI-based methods suffer from low key generation rate and predictable channel attack~\cite{mathur2008radio,jana2009effectiveness}. CSI-based systems can improve key generation rate greatly and are robust to predictable channel attack~\cite{xi2016instant}. However, the major limitation is it requires special toolkits to extract CSI from wireless card, and currently CSI can only be extracted from a limited number of chipsets such as Intel's 5300 NIC~\cite{halperin2011tool,schulz2018shadow}.

To overcome this issue, many recent designs appear for mobile platforms by utilizing the on-board sensory data, such as acoustic data, motion sensor data, bio-sensor data, etc. However, through our study, we find that they mainly trade the sensor's availability for other two further restrictions --- either the system works in a very short pairing distance, e.g., 1.25~cm in Proximate~\cite{mathur2011proximate}, 5cm in TDS~\cite{xi2016instant}, or in a limited set of pre-defined contexts, e.g., in certain environments, when users are performing certain activities~\cite{xu2016walkie,shen2018shake}, etc. The short pairing distance greatly limits the application scenarios of such systems. For example, they cannot work when people need to keep social distance (usually >1.5~\meter) in pandemic. Many other approaches may further suffer a long authentication delay~\cite{han2018you}, need expensive software to support (e.g., public key)~\cite{xie2018genewave}, or require extra hardware (e.g., bio-sensors)~\cite{roeschlin2018device,wang2016touch}. In this paper, we thus aim to investigate whether we can still achieve a fast and ubiquitous key agreement system design that can pair two mobile devices beyond social distance, yet without using extra sensors not available on mobile devices? 

\begin{figure*}[!t]
	\centering
	\subfigure[Reciprocity]{
		\includegraphics[width=1.8in]{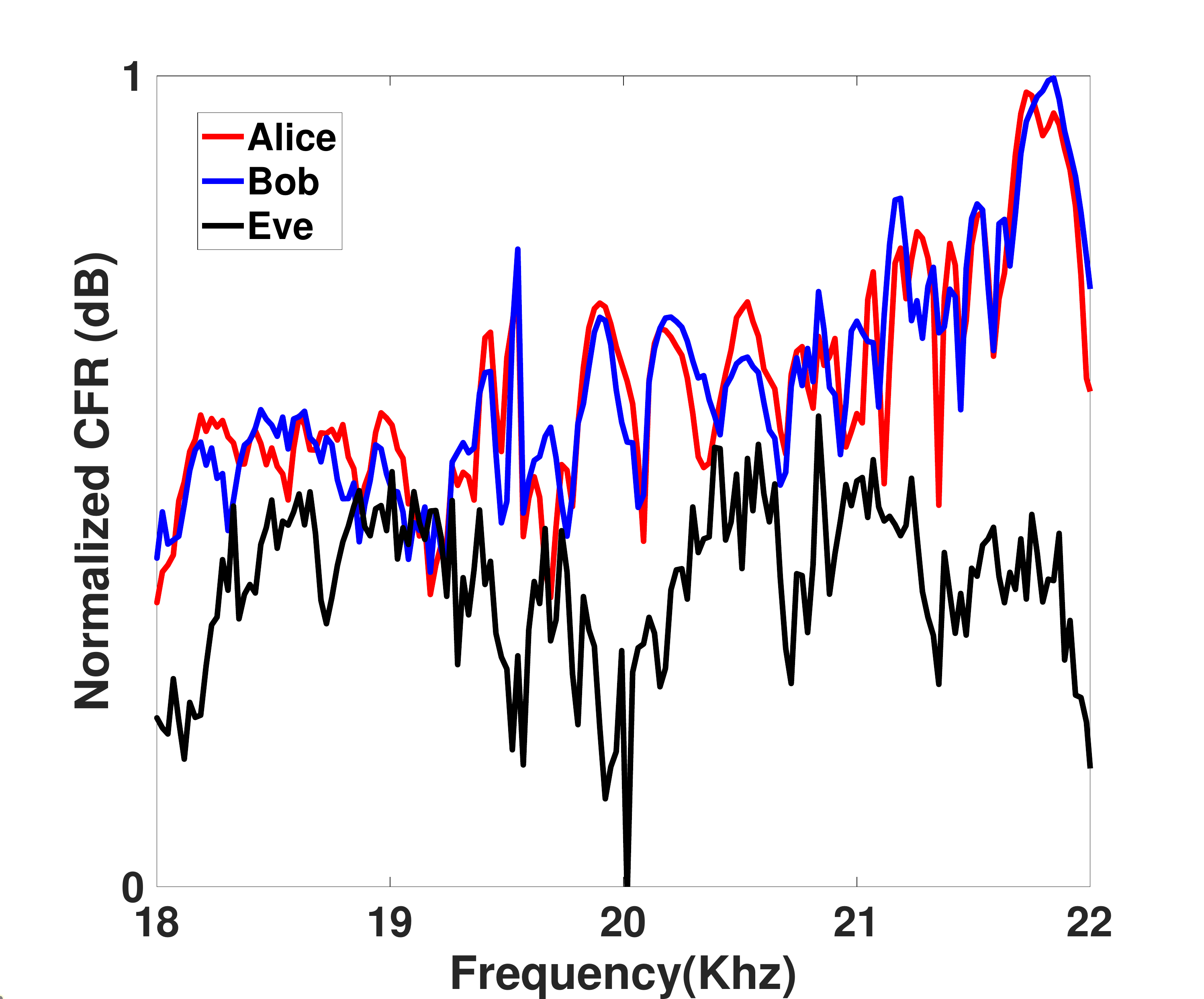}
		\label{fig:reciprocity}
		}
		\vspace{-0.1in}
	\subfigure[Temporal variation]{
		\includegraphics[width=1.8in]{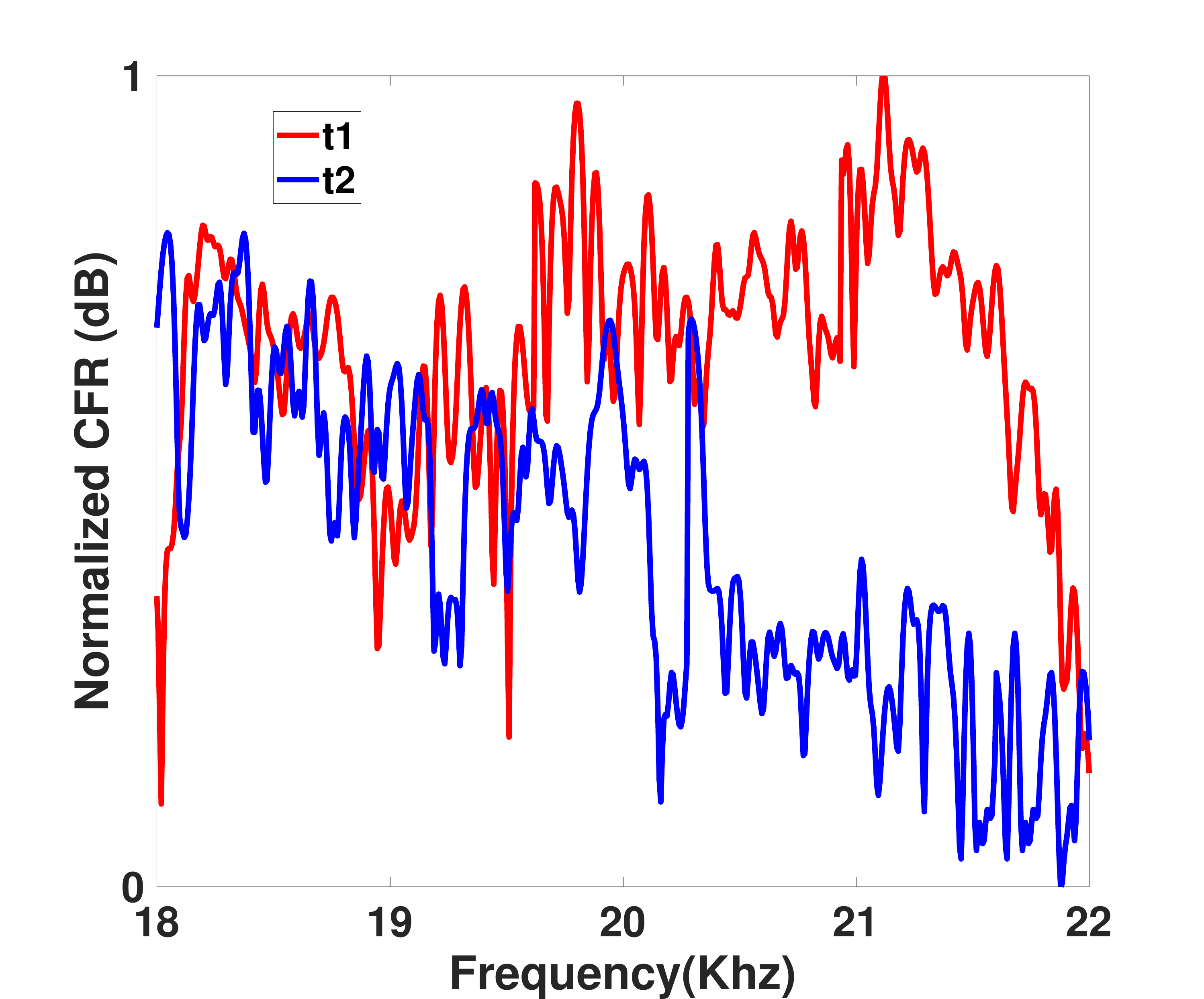}
		\label{fig:temporalvariation}
		}
		\subfigure[Spatial decorrelation]{
		\includegraphics[width=1.8in]{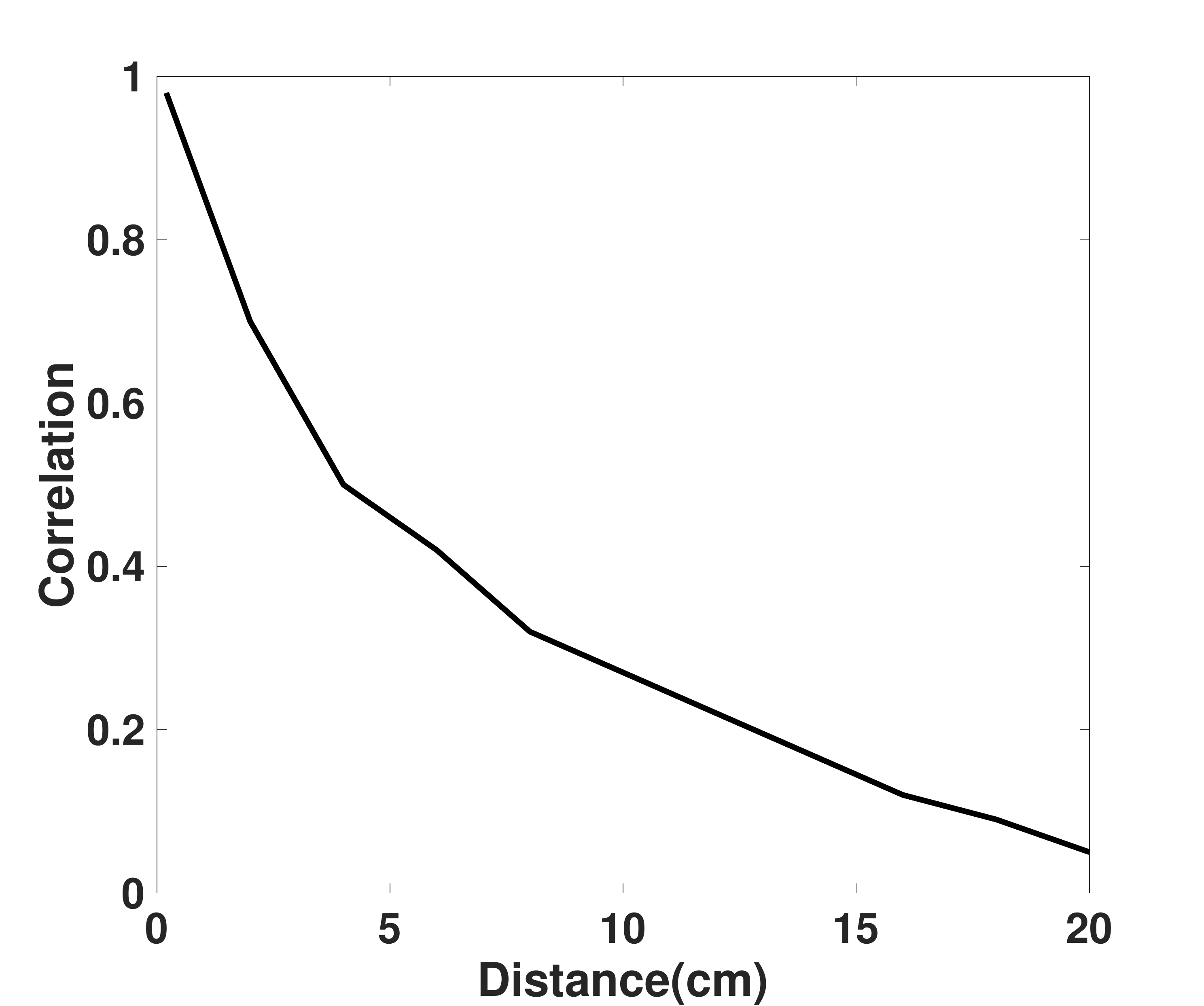}
		\label{fig:spacedecorrelation}
		}
		\vspace{-0.1in}
		\caption{Feasibility Study.}
		\label{fig:feasibility}
		\vspace{-0.1in}
\end{figure*}

\subsection{Our Approach and Design Challenges}

We find that the acoustic signals have such a great potential, inspired by the success from previous radio signal based designs. Acoustic wave, as a form of wave, possesses many properties that radio wave has. In particular, we exploit to leverage \textit{acoustic channel reciprocity} to generate keys. One primary advantage of using acoustic signal is that it is not dependent on special hardware as most mobile devices are equipped with microphone and speaker. Our preliminary study also validates that the acoustic channel indeed holds channel reciprocity, as well as temporal variation and spatial decorrelation, which could serve as the basics of key generation. However, because of the limited acoustic channel bandwidth and the location offset between speaker and microphone, a number of challenges need to be addressed to design an efficient and robust key agreement protocol based on the acoustic signal. 

\begin{enumerate}[leftmargin=*, labelindent=0pt]
    \item Firstly, how to achieve high bit generation rate through narrow band acoustic channels. To this end, we need to extract fine-grained acoustic channel information. Unfortunately, different from wireless card, microphone is not designed to provide channel information such as RSSI and CSI. To extract fine-grained channel information, we design an effective transmitting scheme that uses the inaudible acoustic signal to modulate Orthogonal Frequency-Division Multiplexing (OFDM) symbols. Although applying OFDM modulation in acoustic signals is proposed in FingerIO~\cite{nandakumar2016fingerio}, the use of OFDM in a key generation system can provide more acoustic channel information that can be used to generate keys. As a result, the key generation rate is significantly improved as demonstrated in our evaluation.
    
    \item Secondly, how to further improve the entropy of the extracted key. Existing quantization methods usually use a threshold for the binary encoding~\cite{jana2009effectiveness}. However, these methods may produce repeated bit strings which reduce the entropy of the generated keys. Additionally, these keys are used directly to generate the final key, which leaves opportunities for powerful attackers to obtain raw information by reverse engineering. To tackle this problem, we first apply a novel Bloom filter-based technology to protect the keys against reverse engineering attack. Then, we leverage Karhunen-Loeve Transform (KLT) to remove the redundant information and enhance randomness. 
    
    \item Thirdly, the microphone and speaker are not located at the same location in mobile devices. Hence, the transmitted signal and received signal will experience slightly different channels. Moreover, due to hardware diversity and manufacture imperfection~\cite{zhou2014acoustic}, different microphones/speakers attenuate some frequencies selectively which will further cause more errors. To address this challenge, we optimise a novel compressive sensing (CS) based reconciliation mechanism.  The discrepancies between two initial keys can be corrected with the help of powerful $\ell_1$ optimization. In particular, \SystemName can achieve high matching rate even for different types of IoT devices.
\end{enumerate}
Although several recent works have exploited acoustic signal to pair mobile devices~\cite{han2018you,xie2018genewave,lu2019free,bala2020phy}, our system shows significant performance improvement. We make the following contributions in this paper:
\begin{itemize}[leftmargin=*, labelindent=0pt]
    \item \textbf{System Design.} We propose \SystemName, an inaudible acoustic signal-based key agreement protocol for mobile devices. Based on acoustic channel reciprocity, \SystemName utilizes the channel frequency response of OFDM symbols to generate keys. \SystemName employs several novel approaches to significantly improve the system performance. Particularly, we propose an optimisation algorithm to improve the performance of a state-of-the-art reconciliation method.
    
    \item \textbf{System Implementation.} To demonstrate the feasibility, we implement the prototype of \SystemName on both powerful devices (smartphone) and resource-limited devices (Arduino Uno board). Evaluation results show that \SystemName incurs low system cost and can run efficiently on these IoT devices. We also demonstrate that it is more energy efficient than public key cryptography and authenticated D-H protocol on IoT devices.
    
    \item \textbf{System Evaluation.} We conduct extensive experiments in different real environments.  Compared to state-of-the-art works, \SystemName improves key generation rate by 3 times, extends pairing distance by 3.2 times, reduces information reconciliation counts by 2.5 times. 
    
    \item \textbf{Security Analysis.} Extensive analysis shows that \SystemName is robust to a number of malicious attacks, such as eavesdropping attack, imitating attack and predictable channel attack.
\end{itemize}

The rest of the paper is organized as follows. We present preliminary study results in Sec.~\ref{sec:preliminarystudy}, followed by a description of the system model in Sec.~\ref{sec:systemmodel}. Then we specify the system design in Sec.~\ref{sec:protocol}. We evaluate the performance of \SystemName in Sec.~\ref{sec:evaluation} and analyze its security against attacks in Sec.~\ref{sec:securityanalysis}. Finally, Sec.~\ref{sec:relatedwork} discusses related works, and Sec.~\ref{sec:conclusion} concludes the paper.

\section{Feasibility Study}
\label{sec:preliminarystudy}
We first conduct preliminary study to verify whether acoustic channel hold channel reciprocity, temporal variation and spatial decorrrelation which serve as the basis for key generation.


\textbf{Channel reciprocity.} Channel reciprocity means the channel characteristics (gains, phase shifts, and delays) measured between two devices by exchanging a pair of probe packets within channel coherence time will be very close~\cite{jana2009effectiveness}. For wireless channel, the widely used channel characteristics include RSSI~\cite{jana2009effectiveness,liu2012collaborative} and CSI~\cite{liu2013fast,xi2014keep,xi2016instant}. Unfortunately, the microphone cannot report such channel information. Recent studies use channel taps~\footnote{channel taps are the aggregate of delays caused by multi-path effect~\cite{tse2005fundamentals}.}~\cite{lu2019free} and sound pressure~\cite{bala2020phy} as acoustic channel characteristics. However, we find that they can only provide coarse-grained channel measurements. 

In this paper, we use channel frequency response (CFR), which can provide fine-grained acoustic channel information. CFR means the response of a channel at different frequencies. The multipath fading affects different frequencies across the channel to different degrees giving rise to frequency selective channel~\cite{tse2005fundamentals}. Moreover, if one of the devices is moving such as shaken by the user, it will also cause channel variations due to Doppler effect. Therefore, the received signal will have different responses at different frequencies. In other words, the randomness in acoustic channel responses can be used to generate keys just like radio channel characteristics.

To validate the channel reciprocity, Alice and Bob exchange a number of acoustic signals while Eve (20~cm away) is eavesdropping the acoustic signal between Alice and Bob. Fig.~\ref{fig:reciprocity} plots the CFR of Alice, Bob and Eve. We can see that the CFR of the legitimate devices are close to each other while Eve has different channel responses (though there are some similarities in certain frequencies). However, we also find the CFRs of Alice and Bob are not exactly the same. First, the signals transmitted by Alice and Bob are not transmitted exactly along the same path because the speaker and microphone  are located in different locations on the smartphone. This is different from the wireless devices transmission, where the transceivers could always use one specific antenna. Second, the hardware frequency selectivity from imperfect hardware such as speaker~\cite{zhou2014acoustic}, can also result in the CFR difference. Therefore, we can see different strengthened and weakened power amplitudes at each frequency.

\textbf{Temporal variation.} We set up two mobile phones---namely, Alice and Bob--- in a laboratory with a distance of 1~m. Alice keeps sending inaudible acoustic signals whose frequency varies from 18~kHz to 22~kHz, while Bob is listening to the acoustic channel. Fig.~\ref{fig:temporalvariation} shows the channel frequency response at two different times (t1 and t2 in Fig.~\ref{fig:temporalvariation}) when the user shakes one of the devices. We can see the acoustic channel response at different time instances are different. When the user is shaking device randomly the acoustic channel between Alice and Bob changes rapidly. If the environment changes such as a person walks by, the acoustic channel will also change due to multi-path and Doppler effect just like the radio channel.

\textbf{Spatial decorrelation.} To validate spatial decorrelation, we vary the distance between Eve and Bob from 1~cm to 100~cm. As shown in Fig.~\ref{fig:spacedecorrelation}, the correlation between Eve's and Bob's channel responses decreases rapidly as the distance increases.  According to radio propagation theory~\cite{tse2005fundamentals}, the channel will be statistically uncorrelated if two devices are separated by half wavelength away. If we use 22~kHz frequency audio signal, the wavelength $\lambda$ is 1.7~cm. Therefore, if Eve is located away from a legitimate device by $\lambda = 0.85$~cm, it will have different channel measurements.
In practice, however,  this distance is set as at least multiple wavelengths to alleviate a poor multipath scattering environment or interference and enhance security~\cite{edman2011passive}. From Fig.~\ref{fig:spacedecorrelation}, we can see that the correlation drops below 0.4 when the distance between Eve and legitimate device is greater than 10~cm. Hence, the secure distance is 10~cm in this paper, and we assume any attacker entering this range can be easily spotted by Alice or Bob.
\begin{figure}
	\centering
		\includegraphics[width=2.7in]{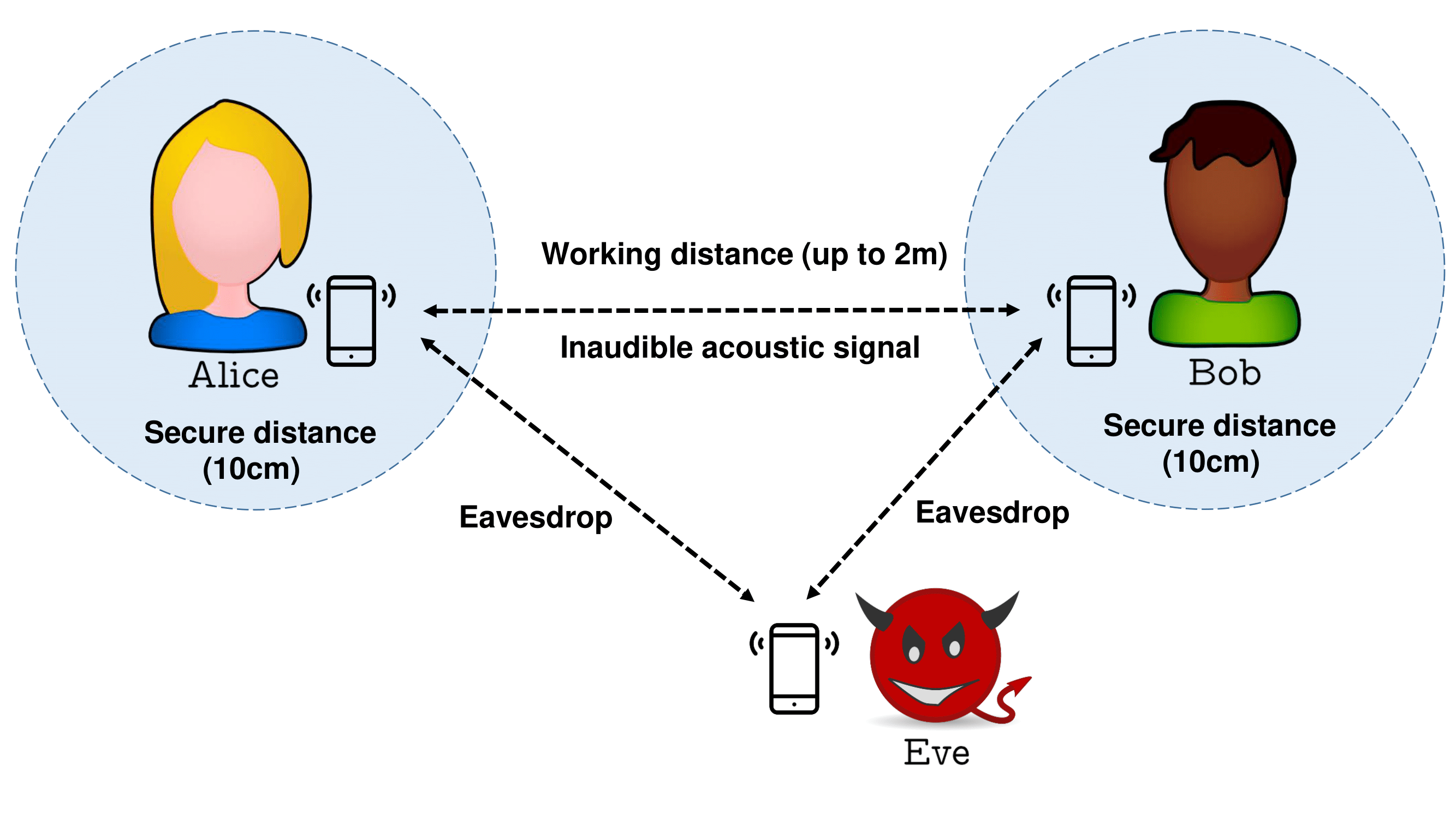}
		\vspace{-0.15in}
	\caption{System model}
	\vspace{-0.16in}
	\label{fig:Systemmodel}
\end{figure}
\section{System Model}
\label{sec:systemmodel}
Fig.~\ref{fig:Systemmodel} illustrates the system model in this paper. We assume that two mobile devices, namely Alice and Bob, intend to generate the same key to secure their communication. Both devices are equipped with a speaker and a microphone. They have no prior shared secrets except that they have \SystemName installed. We assume that an adversary device Eve is located beyond a safe distance (10~cm in \SystemName) to the legitimate devices. If Eve moves within the safe distance, it can be easily spotted by the users as noted in~\cite{xie2018genewave}. One application scenario can be illustrated as follows. Suppose in a meeting, Alice and Bob who meet for the first time want to exchange their name card safely. But to keep social distance, they cannot establish a secure communication channel via existing approaches, such as Touch-and-Guard~\cite{wang2016touch}, shake-n-shack~\cite{shen2018shake} or NFC. By using \SystemName, they can simply shake their device (e.g., smart watch) or perform a random gesture near the device for a short while. A secure communication is then established between them even they are 1-2~m away from each other. If there are sufficient randomness such as moving subjects/objects around users, they do not even need to take any actions (see our demo~\footnote{\url{https://www.youtube.com/watch?v=V8JSgOhairM}}).

We assume that Eve has the full knowledge of the key agreement protocol and can eavesdrop, inject, and replay messages. However, like many previous key generation studies~\cite{xu2016walkie,xi2016instant,xi2014keep,lu2019free,xie2018genewave}, although Eve can inject messages to the public wireless channel, we assume the goal of Eve is to intercept the secret key rather than jamming their communications (i.e., DOS attack). In fact, the DOS attack against \SystemName can be performed by jamming inaudible acoustic signals into the environment. We can use the methods in the literature~\cite{walters2007wireless} to detect such attack. So we consider three types of attacks that are commonly used in related work~\cite{xi2014keep,lu2019free}.
\begin{itemize}
\item Eavesdropping attack: Eve eavesdrops all the messages transmitted in the public channel to extract the same key.
\item Imitating attack: After Alice or Bob finish key extraction, Eve approaches the same site with the aim of generating the same key as a legitimate user. For example, Eve can first observe how Alice or Bob uses devices, e.g. the way of moving or shaking smartphones, then try to imitate his/her use pattern and generate the same key.
\item Predictable channel attack: Eve can deliberately move around to generate desired or predictable changes in the channel between Alice and Bob.

\end{itemize} 

\begin{figure}[!t]
	\centering
\includegraphics[width=3.5in]{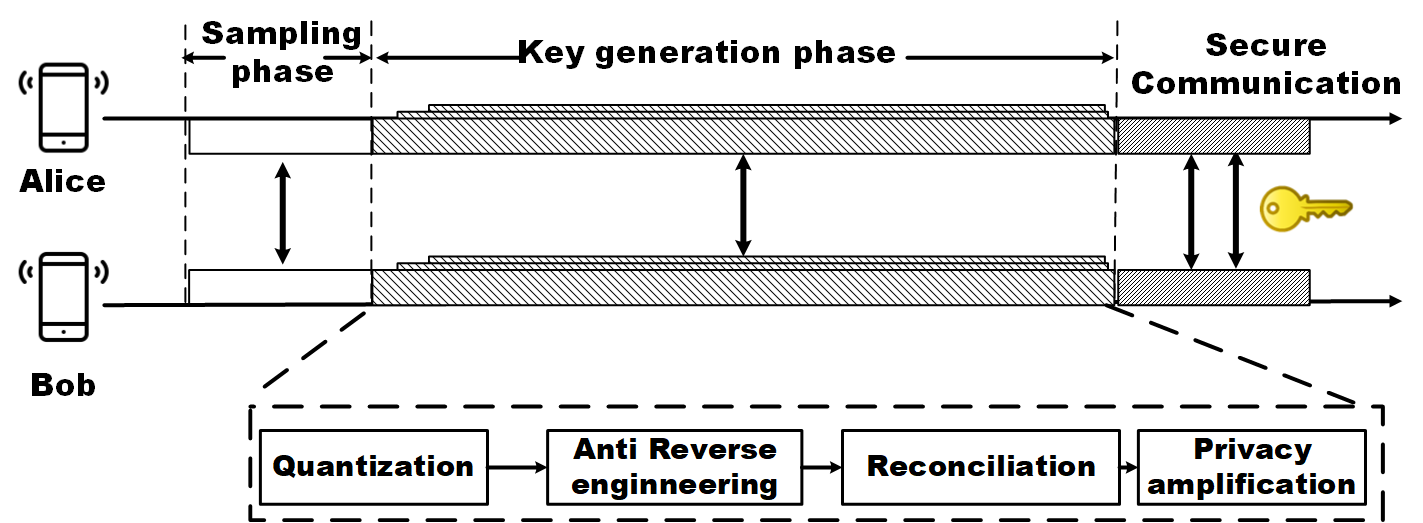}
		\vspace{-0.15in}
	\caption{System Flowchart.}
	\label{fig:architecture}
	\vspace{-0.2in}
\end{figure}
\vspace{-0.1in}
\section{System Design}
\label{sec:protocol}
Fig.~\ref{fig:architecture} shows the work-flow of \SystemName. Suppose Alice and Bob are two devices that want to generate a secret key. Firstly, they exchange a number of inaudible acoustic frames and calculate the CFR. Then both devices follow the steps in Fig.~\ref{fig:architecture} to generate the same cryptographic key. Finally, they use the extracted key to secure their communication.
\subsection{Transmitting signal design}
\label{sec:transmitsignal}
In sampling phase, both Alice and Bob exchange a number of acoustic frames by transmitting via speaker and receiving via microphone to obtain channel measurements. \SystemName uses inaudible frequency band from 18~kHz to 22~kHz for not disturbing users. 

To obtain fine-grained channel information, we apply OFDM technology on an acoustic signal based on the method in FingerIO~\cite{nandakumar2016fingerio}. Specifically, we divide the 18-22~kHz frequency band into 64 subcarriers so that the width of each subcarrier is 62.5Hz. The transmitting time-domain samples can be obtained by performing inverse Fast Fourier transform (IFFT) on the transmitted data, and the receiver can reconstruct the raw data bits by a Fast Fourier Transform (FFT). A speaker will transmit the vectors with 64 real values from the OFDM symbol construction. Another advantage of using OFDM technology is that both devices can probe channel within channel coherence time without explicitly synchronising two mobile devices. In fact, it is impractical to assume two mobile devices are synchronised when they first meet each other. We use the first $S_{suf}$ of these values to form a cyclic suffix that is appended to the end of OFDM symbol. The cyclic suffix is used to accurately estimate the beginning of the OFDM symbol. Even Alice and Bob are not synchronized, they can still locate the beginning of the received symbol by calculating the correlation between the received signal and known transmitting signal (see~\cite{nandakumar2016fingerio} for more details).
\begin{figure*}[!ht]
	\centering
	\hspace{-0.3in}
	\subfigure[Impact of window size]{
		\includegraphics[width=2.5in]{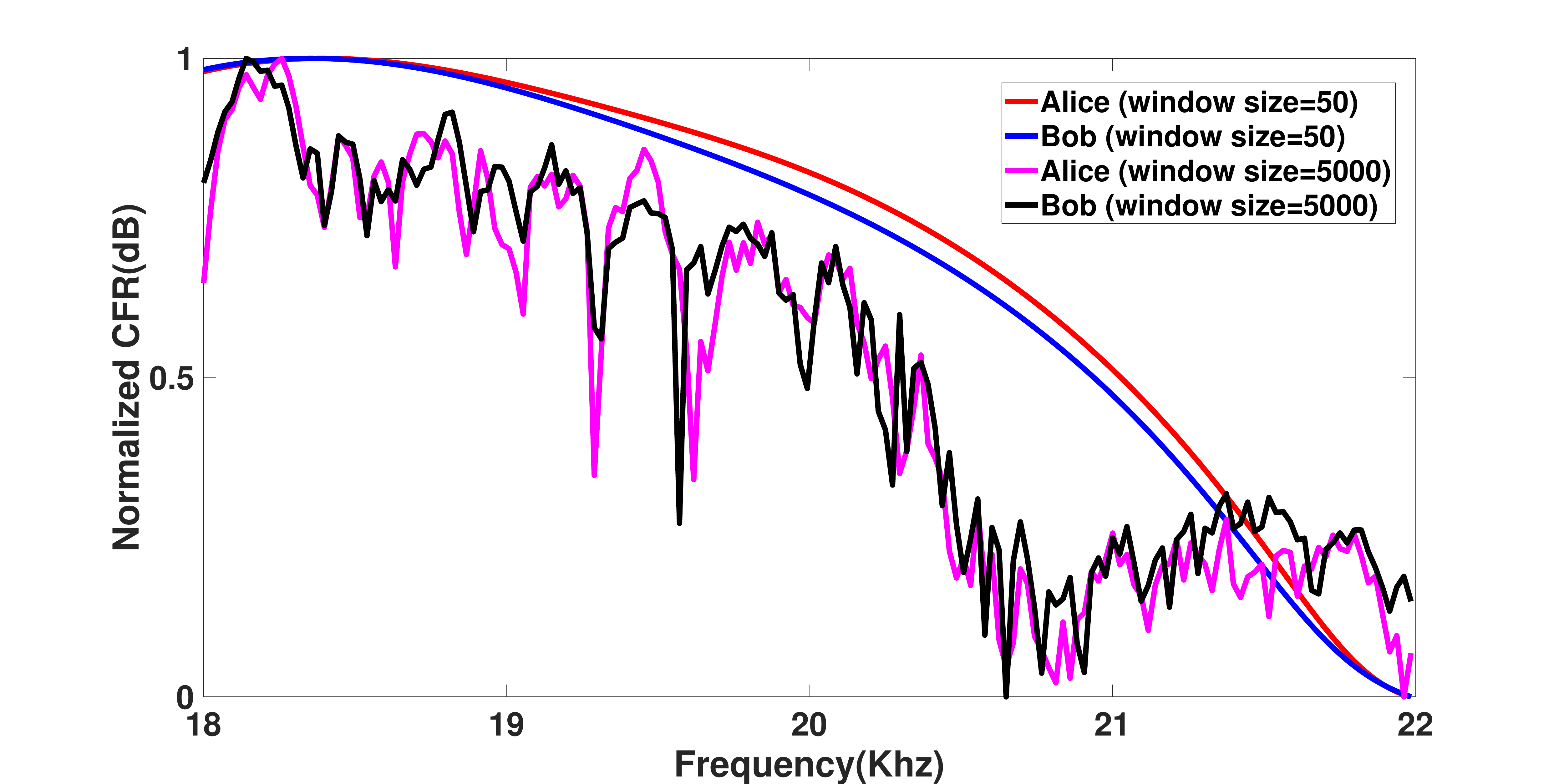}
		\label{fig:impactwindowsize}}
		\hspace{-0.3in}
		\subfigure[An example of quantization]{
		\includegraphics[width=2.5in]{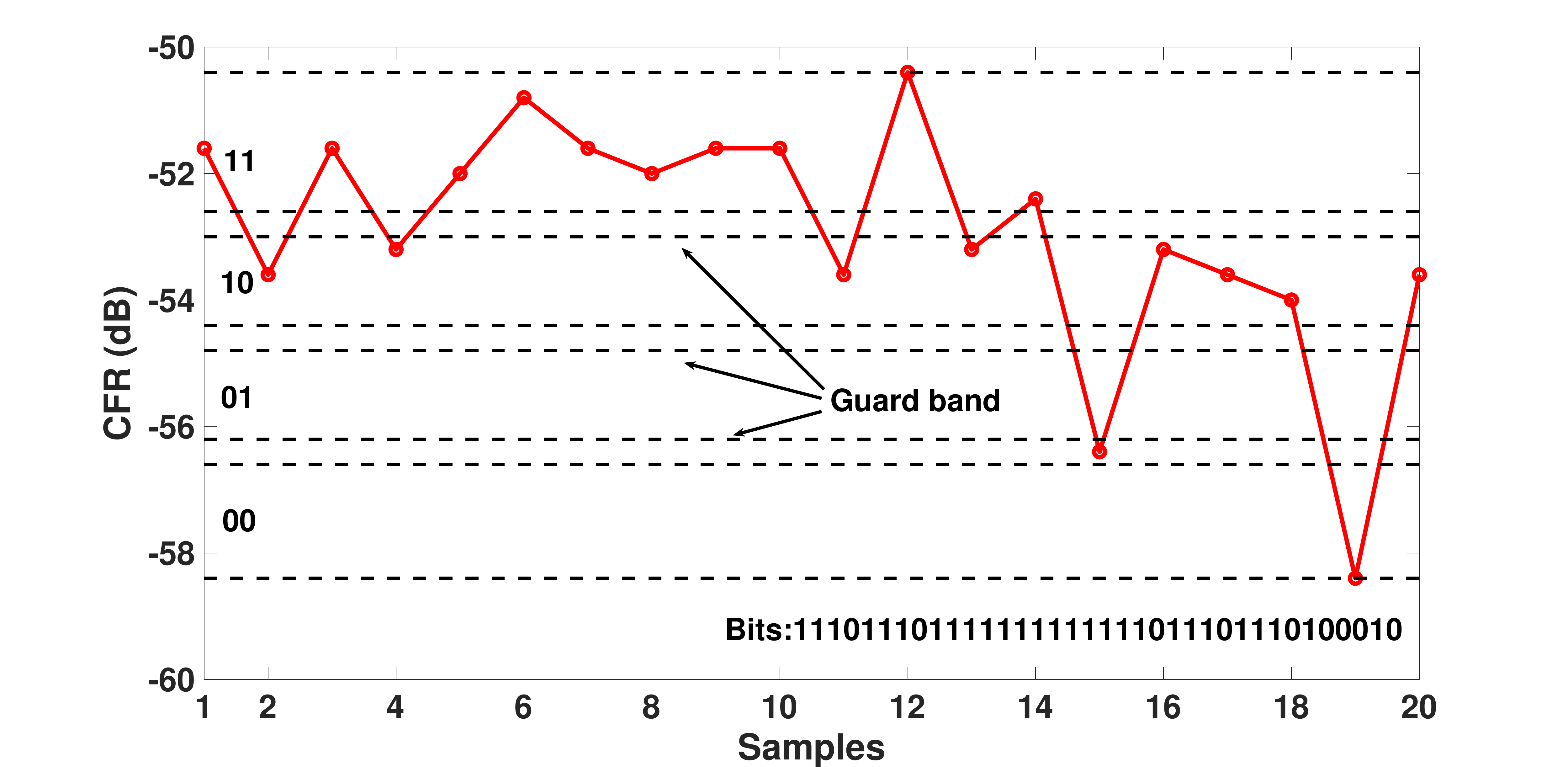}
		\label{fig:quantization}}
		\hspace{-0.3in}
		\subfigure[Projecting the Bits into Bloom filter space]{
		\includegraphics[width=2.5in]{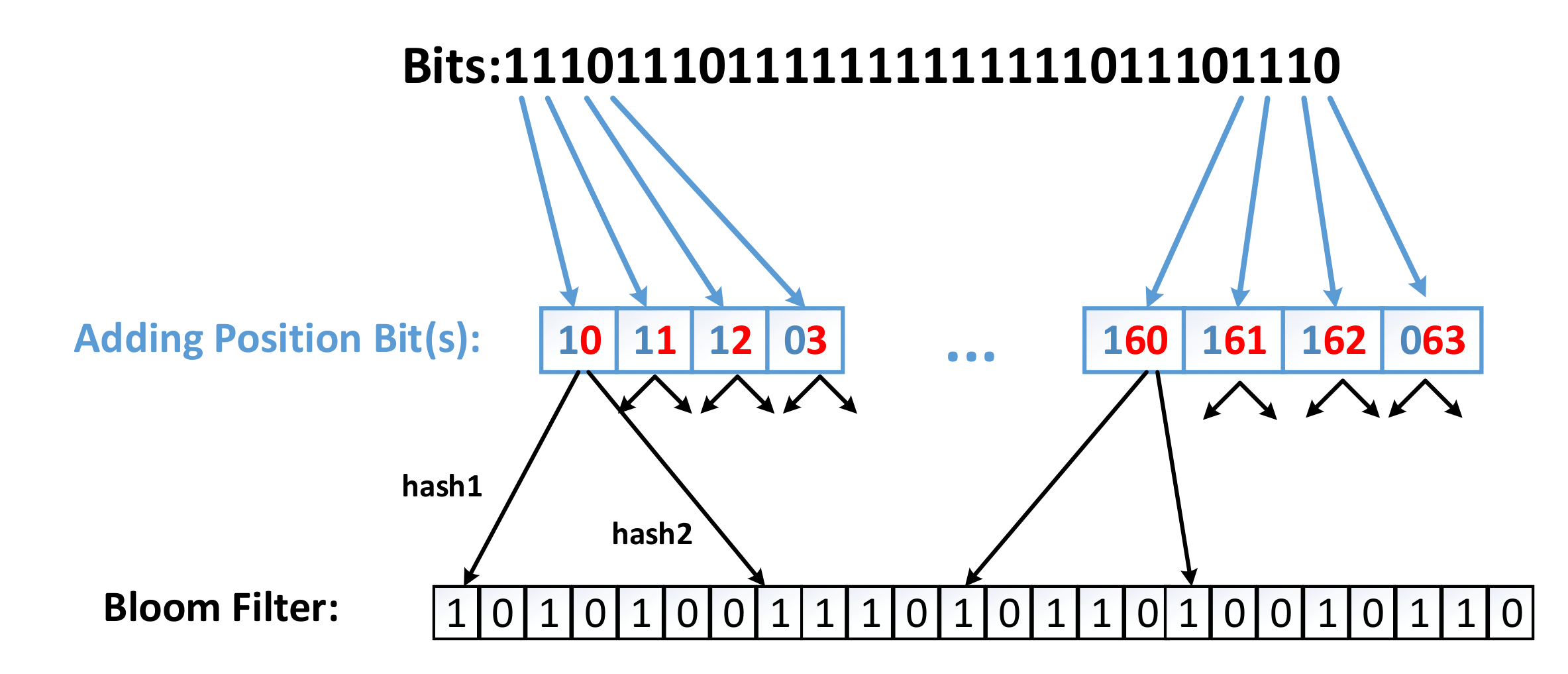}
		\hspace{-0.3in}
	\label{fig:bfdesign}}
	\vspace{-0.1in}
	\caption{Illustration of quantization process.}
	\vspace{-0.15in}
\end{figure*}

The length of the transmitting signal and the transmitting interval is important for the following two reasons. 1) We need to make sure Alice and Bob obtain the channel estimation within coherence time, so their CFRs are highly correlated; 2) the transmitting interval should be larger than the coherence time. Otherwise, the consecutive CFRs will be correlated and the randomness of the key will be reduced. In theory, the rate at which the channel varies is represented by Doppler frequency ($f_{d}$) and the duration when the channel is stable which is denoted by channel coherence time ($T_{c}$).  Coherence time is the time domain dual of Doppler spread and is used to characterize the time varying nature of the channel frequency.
Suppose the moving speed of the subject or object is $v$, the channel frequency is $f$, and the speed of acoustic signal is $c$ ($340 m/s$),  then the maximum Doppler frequency is $f_{d} = \frac{v\cdot f}{c}$\cite{rappaport1996wireless}. Practically, the channel coherence time with respect to the maximum Doppler frequency shift is $T_{c}=\sqrt{\frac{9}{16\pi f^{2}_{d}}}$ according to \cite{rappaport1996wireless}. 

The acoustic signal used in \SystemName is from 18~kHz to 22~kHz, and the speed of common human motions varies from 0.1--2.7~m/s~\cite{zhang2003measurement}. Therefore, the coherence time lies in the range of 2~ms to 53~ms.  In \SystemName, we set the length of the cyclic suffix $S_{suf}$ to 26. Therefore, the transmitted symbol contains 90 samples. Given 48~kHz sampling rate, the transmitter takes 1.9~ms to transmit these 90 samples, which is shorter than the minimum coherence time. In terms of transmitting interval, Alice and Bob exchange acoustic signals every 100~ms, which is longer than the maximum coherence time. 

\subsection{Quantization}
\subsubsection{Multiple-bit quantizer}
\label{subsubsec:multiplequantizer}
Upon receiving the signal from Bob, Alice applies a Butterworth band-pass filter (18--22~kHz) to filter out the environmental noises and calculate the CFR using the method in~\cite{xie2018genewave}. When calculating CFR, the window length plays an important role: if it is too small, there is not much entropy; however, if it is too large, there will be more mismatches (Fig.~\ref{fig:impactwindowsize}). We empirically use a Hamming window whose size is 2000 in \SystemName.

Then the CFR is quantized to binary bits (0s and 1s) by multiple-bit quantization technique proposed in~\cite{jana2009effectiveness}. To be specific, we first divide the CFR measurements into several windows with no overlap (window size $W=20$). Then for each window, we divide the samples into multiple quantization levels. Each level in the quantization is assigned an \textit{n}--bit code. For example, if $n=2$, then each sample will be converted into 2 bits. We also insert guard band between different levels to mitigate the effect of mismatches. We use $\alpha \in [0,1]$ to represent the ratio of guard band to data. The larger the $\alpha$ is, the more mismatches are discarded. However, it is possible that the length of the keys generated by Alice and Bob are different. To solve this issue, we exchange the indices of samples that are used to generate keys and only reserve keys generated at the common indices. Fig.~\ref{fig:quantization} shows how to convert a window of 20 samples into bits. After quantization, we assume the keys generated by Alice and Bob are denoted by $K^{'}_{Alice}$ and $K^{'}_{Bob}$.
\subsubsection{Anti Reverse Engineering}
\label{subsub:privacypreserving}

Most previous work on key generation utilize the quantized bits directly to get the final secret key. However, Eve can perform attacks to derive the key from data collected by herself and the data transferred between Alice and Bob. The Bloom filter has been used as part of the encoding and perturbation methods in many privacy-preserving applications~\cite{erlingsson2014rappor,xue2020sequence}. Unfortunately, traditional Bloom filter projections cannot remain the order information (without additional process). In other words, the mismatches between the two input key strings and output key strings may be different. In \SystemName, we use a special designed Bloom filter data structure considering sequence/order information, which can help project the key into Bloom filter. The purpose of utilizing the adapted Bloom filter is to keep the key distance information while in a non-plaintext format.

The detailed process is presented by the Fig.~\ref{fig:bfdesign}. Take a 64-bit key as an example, each single bit position information is conveyed by an addition bit(s) before Bloom filter hash-mapping. Afterwards, two hash functions are used to hash-map each element in the adding-position-bits to the Bloom filtered space. The Bloom filter hash-mapping will only turn the `0' into `1' based on the hash value (calculation) (see~\cite{bloom1970space} for the original rationale). As a result, each `1' in Bloom filter refers to a bit (`0' or `1') at a certain position of the original key `uniquely'. Most importantly, this adapted Bloom filter data structure can also hold the Jaccard distance between the raw data bits and the projected Bloom filter data bits. That is to say, suppose $K_{Alice}$ and $K_{Bob}$ represent the Bloom filter output of $K^{'}_{Alice}$ and $K^{'}_{Bob}$, if there are $N_{mis}$ mismatches between $K^{'}_{Alice}$ and $K^{'}_{Bob}$, then there will be also $N_{mis}$ mismatches between $K_{Alice}$ and $K_{Bob}$.  The proof of this can be referred from ~\cite{xue2020sequence}. Thus, we can directly use the following information reconciliation approach on $K_{Alice}$ and $K_{Bob}$ because they preserve the similarity information between $K^{'}_{Alice}$ and $K^{'}_{Bob}$. Note that although the Boom filter is an irreversible one-way function, if the input key's length is too short, Eve can still get the Bloom filter's output, such as through brute-force attack. Therefore, we need to ensure the entropy of the input of the Bloom filter. In \SystemName, we concatenate the bits generated from each window to a key string and further divide it into consecutive segments where each segment contains 128 bits. Since Eve has no knowledge of the number and location of the incorrect bits, it is computationally infeasible ($2^{128}$ guesses) to obtain $K^{'}_{Alice}$ and $K^{'}_{Bob}$.  The adapted Bloom filter, which only use hash functions and limit temporary storage (to store the Bloom filter results), will not involve much overhead to our entire mobile system.

\vspace{-0.1in}
\subsection{Information Reconciliation}
\label{sub:reconciliation}

Because of noise, we usually get $K_{Alice}\approx K_{Bob}$ instead of exactly identical keys. The purpose of reconciliation is to correct the mismatches between $K_{Alice}$ and $K_{Bob}$. In \SystemName, we optimise a recent developed CS-based reconciliation method~\cite{lin2019h2b} to improve the key agreement rate. To make this paper more self-contained, we first succinctly describe the flow of this method then present our optimisation algorithm.  

The key idea of the reconciliation method~\cite{lin2019h2b} is \textit{the mismatches between Alice and Bob is much less than that of Alice and Eve (i.e., more sparse)}. Suppose the keys after bloom filter are  $K_{Alice} \in R^{N}$ and $K_{Bob} \in R^{N}$, respectively. The sampling matrix is $A \in R^{M \cdot N}$ which obeys the restricted isometry property (RIP)~\cite{Donoho06}. Researches have found a random Bernoulli matrix with equally distributed works well, so we use a random Bernoulli matrix in \SystemName. Then Alice and Bob follow the steps below to correct their mismatches.
\begin{enumerate}[noitemsep,topsep=0pt,leftmargin=*, labelindent=0pt]
\item Firstly, Bob generates a compressed  vector $y_{Bob}=A\cdot K_{Bob} \in R^{M}$ and transmits it to Alice via public channel.
\item Secondly, after receiving $y_{Bob}$, Alice calculates the difference between $y_{Alice}$ and  $y_{Bob}$:
\begin{align}
		   y_{A,B} = y_{Alice}-y_{Bob}&= A (K_{Alice}-K_{Bob}) + e= A \Delta x +e
    \label{eq:reconciliation2} 
\end{align}
where $y_{Alice} = A K_{Alice}$ and $\Delta x$ represents the mismatches between $K_{Alice}$ and $K_{Bob}$ and $e \in R^{M}$ is noise.

\item Finally, Alice apply $\ell_1$ optimization to reconstruct $\Delta x$ from the compressed data $y_{A,B}$~\cite{Donoho06}: 
\begin{equation}
\label{eq:l1}
\argmin_{\Delta x} \| \Delta x \|_1 \quad \text{ subject to } \|y_{A,B} - A \Delta x\|_2 < \epsilon.
\end{equation}

where $\epsilon$ is used to account for noise. Then, Alice can deduce  $K_{Bob}$ by $\bar{K}_{Alice} = K_{Alice} \oplus \Delta x$, and both Alice and Bob agree on the same key $\bar{K}_{Alice} = K_{Bob}$.
\end{enumerate}

When we use the above method, we find that the performance varies a lot. We find out it is because the sampling matrix $A$ is generated randomly. To address this problem, we propose an optimisation problem to improve its performance. According to the theory of compressive sensing~\cite{candesrip}, the random matrix $A$ must meet the following two conditions.
\begin{equation}
M(A)\le\frac{C}{\log N}, \quad s\le\frac{CN}{\log N\cdot\|A\|_2^2}
\label{eq:conditions}
\end{equation}
where $C$ is a constant, $N$ is the length of the key (e.g., the number of columns of $A$), $s$ is the sparsity of the key, and $M(A)$ means the mutual coherence of $A$ which is defined as:
$M(A) = \max_{i < j} \frac{| a_i^T a_j |}{\|a_i\| \|a_j\|}$
where $a_i$ and $a_j$ represent the $i-{th}$ and $j-{th}$ columns of $A$ respectively. In fact, $M(A)$ represents the maximum value of cross-correlation between columns in $A$. If $A$ is generated randomly every time, it is hard to ensure it always has good mutual coherence. That is why the performance varies greatly for different $A$. If we change the second condition above into this form: $\frac{1}{\|A\|_2^2}\ge\frac{s\log N}{CN}$, we can see that if $\|A\|_2^2$ is smaller, it is easier for $A$ to meet the second condition. Therefore, the goal of our proposed optimisation algorithm is to minimise $M(A)$ iteratively. 

As shown in Algorithm~\ref{Alg:1}, we start with finding two columns that have minimum coherence from a searching space $\Omega$. $\Omega$ includes a large finite set of random matrices. Then in each iteration, we choose a vector from $\Omega$ that can minimise the maximum mutual coherence between this vector and those already in $A$. Finally, when we find $N$ such columns, the iteration terminates and we output the optimised matrix $A$ which has the minimum mutual coherence. This optimisation is conducted offline, so it does not incur any computation overhead. Note that although several other works also optimise projection matrix by minimising mutual coherence or row coherence ~\cite{shen2014face,elad2007optimized}, the problems to be solved are different, i.e., they aim to optimise a projection matrix for a certain dictionary to obtain better recovery signal.
\begin{algorithm}[!h]
\small
\caption{Sampling Matrix Optimisation}
\label{Alg:1}
\begin{algorithmic}
\State \textbf{Objective}: Find a matrix A with minimal $M(A)$
\State \textbf{Input}: Search space $\Omega$, the number of columns of $A$ is $N$.
\State \textbf{Initialisation}: traverse $\Omega$ to find two column vectors $\hat{a}_1$ and $\hat{a}_2$ such that their coherence is minimal, $\hat{A}=\{ \hat{a}_1, \hat{a}_2 \}$, $\Omega = \Omega \backslash \{ \hat{a}_1,\hat{a}_2\}$, $i = 2$
\While {$i \leq N$}
\State $\tilde{A}_j=\argmin_{\hat{a}_j\in\Omega}\max_{\hat{a}_k\in\hat{A}}|\hat{a}_j \hat{a}_k|$
\State $\hat{A}=\hat{A}\bigcup\{\tilde{a}_j\}$
\State $\Omega = \Omega \backslash \{ \tilde{a}_j \}$  
\State{$i$++}
\EndWhile
\State \textbf{Output}: optimised matrix $A=\left[a_1,a_2,\cdots,a_m\right]$
\end{algorithmic}
\label{alg:gs} 
\end{algorithm}

The optimised reconciliation method presents several advantages. Firstly, Bob only transmits the compressed key instead of the original key. Even if Eve eavesdrops this message, she cannot reconstruct Bob's key $K_{Bob}$ as will be discussed in Sec.~\ref{sec:securityanalysis}. So it ensures security. Secondly, unlike some conventional reconciliation methods, this approach does not discard errors during reconciliation process. With the powerful $\ell_1$ optimization, this approach can recover more errors than previous reconciliation methods as will be demonstrated in Sec.~\ref{subsec:comparison_reconciliation}. Thus, it can improve key generation rate. Finally, some methods require both Alice and Bob to perform reconciliation to correct the errors between their keys. However, in the CS-based approach, only one device needs to perform reconciliation (e.g., Alice in this example). The users can choose to run $\ell_1$ optimization on the power-rich devices to mitigate the computational burden on resource-limited IoT devices. Thus, this approach can improve energy efficiency as will be demonstrated in Sec~\ref{subsec:implementation}.

As discussed in Sec~\ref{sec:systemmodel}, Eve has the power to modify, insert and replay messages. So she can perform two common attacks during reconciliation process: MITM and replay attack.  Eve can launch MITM by impersonating as Alice or Bob during key generation process to modify or insert her own messages.  To solve this problem, we apply the message authentication code (MAC) method to ensure the integrity and authenticity of the message~\cite{mathur2008radio}. Bob includes an additional MAC message with $y_{Bob}$, so the total message sent to Alice becomes $L_{Bob}=\left\{y_{Bob},MAC(K_{Bob},y_{Bob})\right\}$. After receiving $L_{Bob}$, Alice computes $K_{Alice}^{'}$ by Eq.~\ref{eq:l1} and verifies its identity. If $MAC(K_{Alice}^{''},y_{Bob})\neq MAC(K_{Bob},y_{Bob})$, Alice knows that the message was modified, indicating the presence of Eve. If $MAC(K_{Alice}^{''},y_{Bob})=MAC(K_{Bob},y_{Bob})$, Alice can confirm that this message was indeed originated from Bob. To detect replay attacks, we can adopt some commonly used methods such as nounces, timestamps or tagging each message with a session ID~\cite{malladi2002preventing}.

Although the above methods can prevent MITM and replay attacks, Eve can utilise the eavesdropped $y_{Bob}$ to infer the shared key $y_{Bob}$. The authors in~\cite{lin2019h2b} pointed out two potential vulnerabilities.

\textbf{Vulnerability 1}: Eve can try to recover Bob's key from $y_{Bob}$ by $\ell_1$ optimisation directly. 

\textbf{Vulnerability 2}: Eve may launch the three types of attacks mentioned in Sec.~\ref{sec:systemmodel} to obtain her own channel measurement. Then she can impersonate as a legitimate device and perform the same reconciliation process as Alice by using her channel measurement with the aim of generating the same key.

The authors in~\cite{lin2019h2b} have proved that the CS-based reconciliation approach is perfectly effective and secure if the number of rows of $A$ $M$ meets the following condition:
\begin{equation*}
P < M < Q, \quad P = S_{\Delta A, B} * log(N/S_{\Delta A, B}), \quad Q = min(S_{Bob}, S_{\Delta B, E})
\end{equation*}

where $S_{\Delta A, B}$ is the sparsity of $\Delta x$, $S_{Bob}$ is the sparsity of $K_{Bob}$, $S_{\Delta B,E}$ is the sparsity of difference between $K_{Bob}$ and $K_{Eve}$, respectively. The sparsity here means the number of non-zero values in the corresponding vector, i.e., the number of mismatches. As noted in~\cite{lin2019h2b}, the design for an effective and secure CS-based reconciliation is a problem to find a suitable $M$, with upper bound $Q$ being the secure threshold, and the lower bound $P$ being the effective threshold. We conduct extensive experiments to find a proper range of $M$ in Sec.~\ref{sec:securityanalysis}.
\subsection{Privacy Amplification}
\label{subsec:privacyamplification}
Although multi-level quantization can generate more bits from one sample, it may also generate some duplicated bits as the example in Fig.~\ref{fig:quantization}. Directly using such a key will weaken the security of the system. Note that the Bloom filter-based approach in Sec~\ref{subsub:privacypreserving} can prevent reverse engineering attack but cannot improve entropy. To address this problem, we use  Karhunen-Loeve Transform (KLT) to decorrelate the bit sequence after reconciliation.

Assume the generated key by Alice after reconciliation is $\bar{K}_{Alice}=(k_{1},k_{2},\cdots,k_{L})^{T}$, where $k_{i}$ is the $i$-th bit and $L$ is the length of the key. The first step in KLT is to calculate the auto-correlation matrix $R=E(K K^{T})$. Next, we calculate its eigen value $\lambda_{i}$ and eigen vector $\phi_{i}$ so that $R \phi=  \lambda_{i}\phi_{i} \quad (i=1,2,\cdots,L)$. Note that $R$ is Hermitian, and its eigenvectors ${\bf\phi}_i$ are orthogonal. If we rank eigen values in a descending order, $\lambda_{1} > \lambda_{2} > \cdots >\lambda_{L}$, we can construct an unitary matrix $\Phi$ which diagonalizes $R$: $\Phi = \left[ \phi_1, \cdots, \phi_L \right]$. $\Phi$ is called the KLT matrix and can be used to decorrelate the bit sequences $K$. We choose the largest $S$ eigenvectors to construct $\Phi$, so we can obtain a decorrelated key string by $ K_{Alice}^{''}= \Phi^{T} \bar{K}_{Alice}$. In the same way, Bob can generate a decorrelated key sequence $K_{Bob}^{''}=K_{Alice}^{''}$.


Although reconciliation can improve the reliability of a key generation protocol, it reveals partial information to Eve. Privacy amplification is a common approach to remove the revealed information from the generated secret key sequence. It is usually implemented by the extractor, universal hashing functions, and cryptographic hash functions~\cite{zhang2016key}. In \SystemName, we use the commonly used dual universal hash function~\cite{hayashi2016more} to generate the final key. Finally, the key can be used by encryption algorithms such as AES-128 to secure their communication.
\section{Evaluation}
\label{sec:evaluation}
\subsection{Goals, Metrics and Methodology}
\begin{figure*}[!h]
\hspace{-0.1in}
	\subfigure[]{
		\includegraphics[width=1.25in]{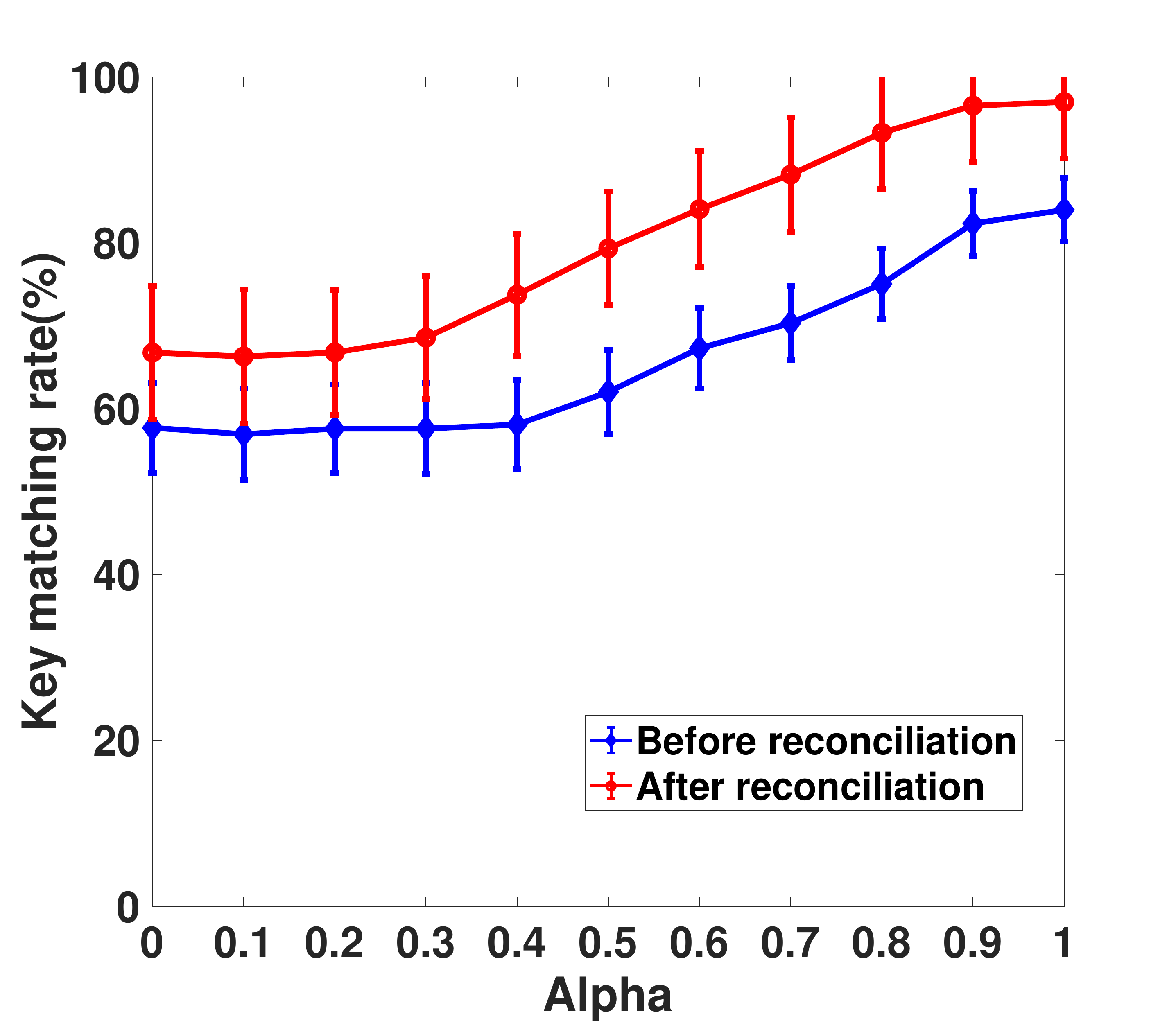}
		\label{fig:key_agreement_rate_alpha}}
		\hspace{-0.1in}
		\subfigure[]{
		\includegraphics[width=1.25in]{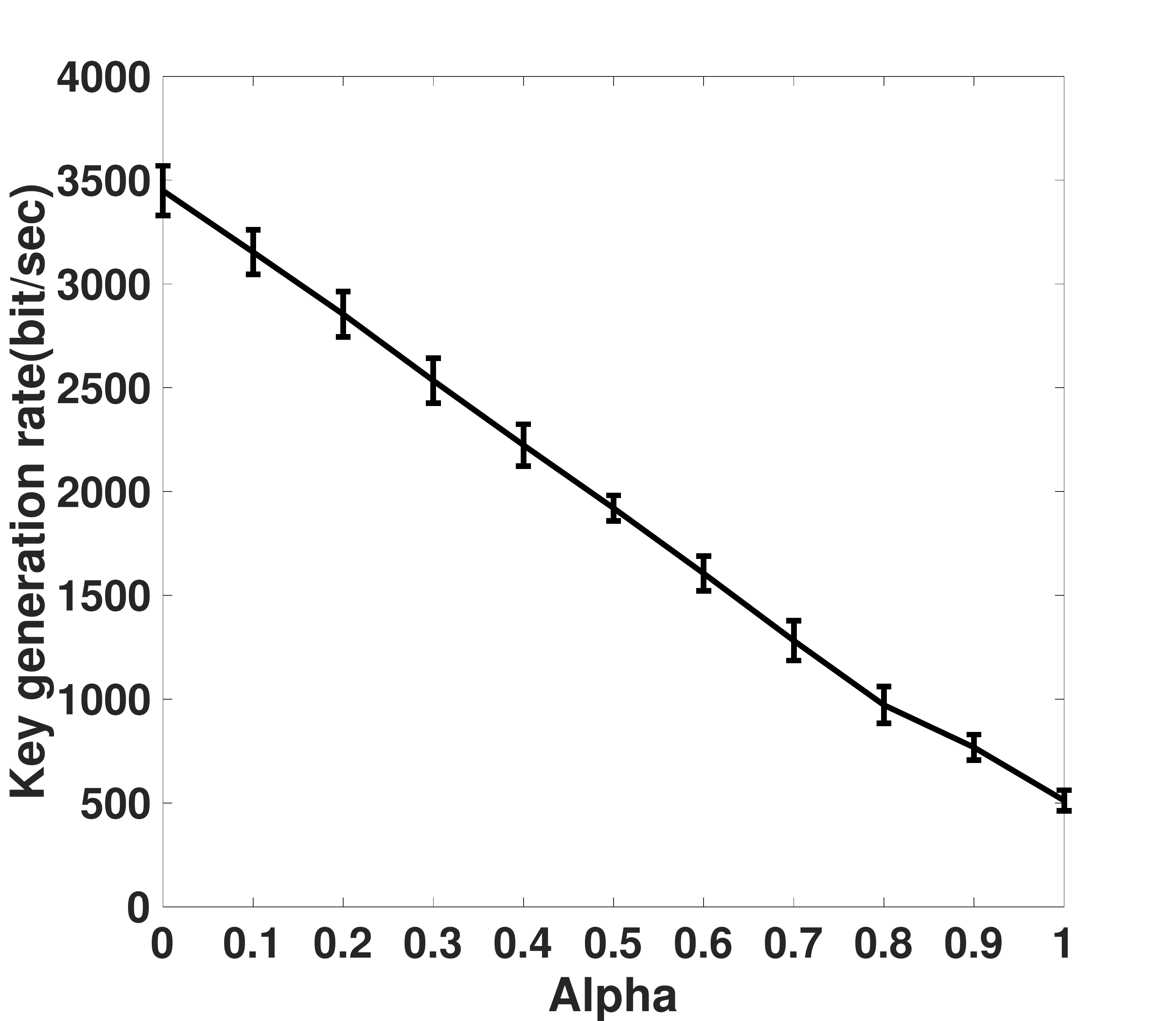}
		\label{fig:key_generation_rate_alpha}}
		\hspace{-0.1in}
		\subfigure[]{
		\includegraphics[width=1.25in]{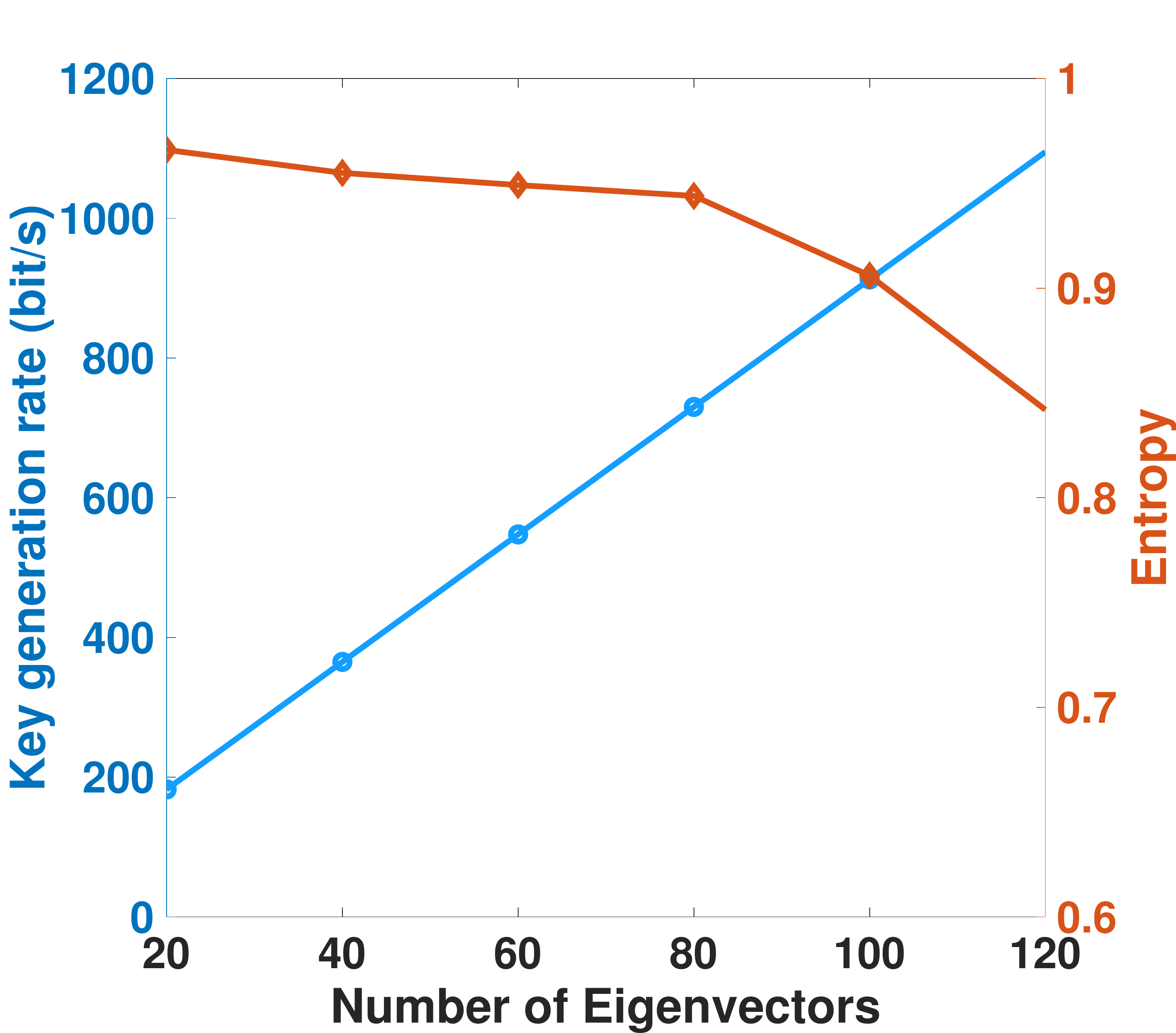}
		\label{fig:Impact_S}}
		\subfigure[]{
		\includegraphics[width=1.3in]{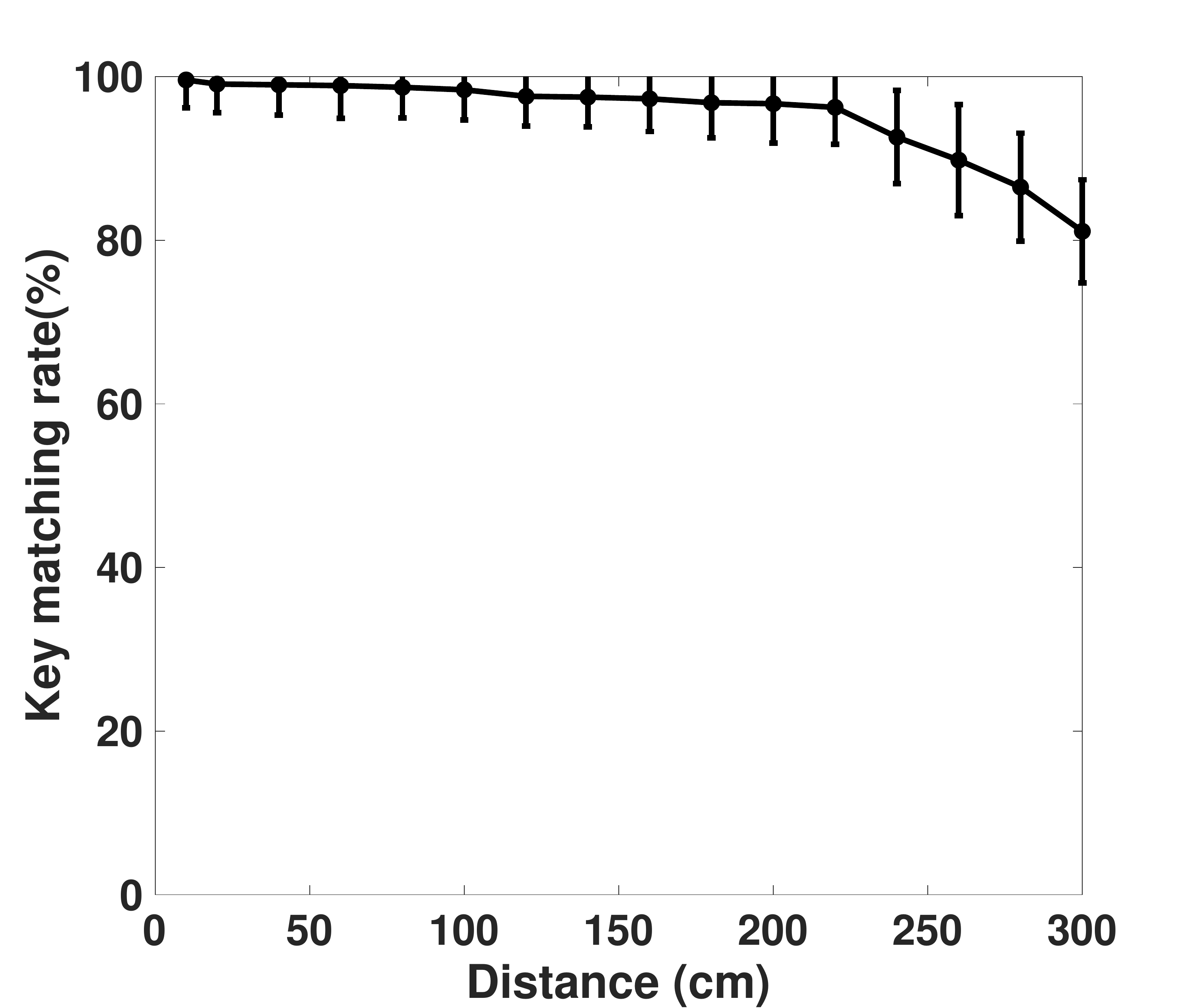}
		\label{fig:key_agreement_rate_distance}}
		\hspace{-0.1in}
		\subfigure[]{
		\includegraphics[width=1.3in]{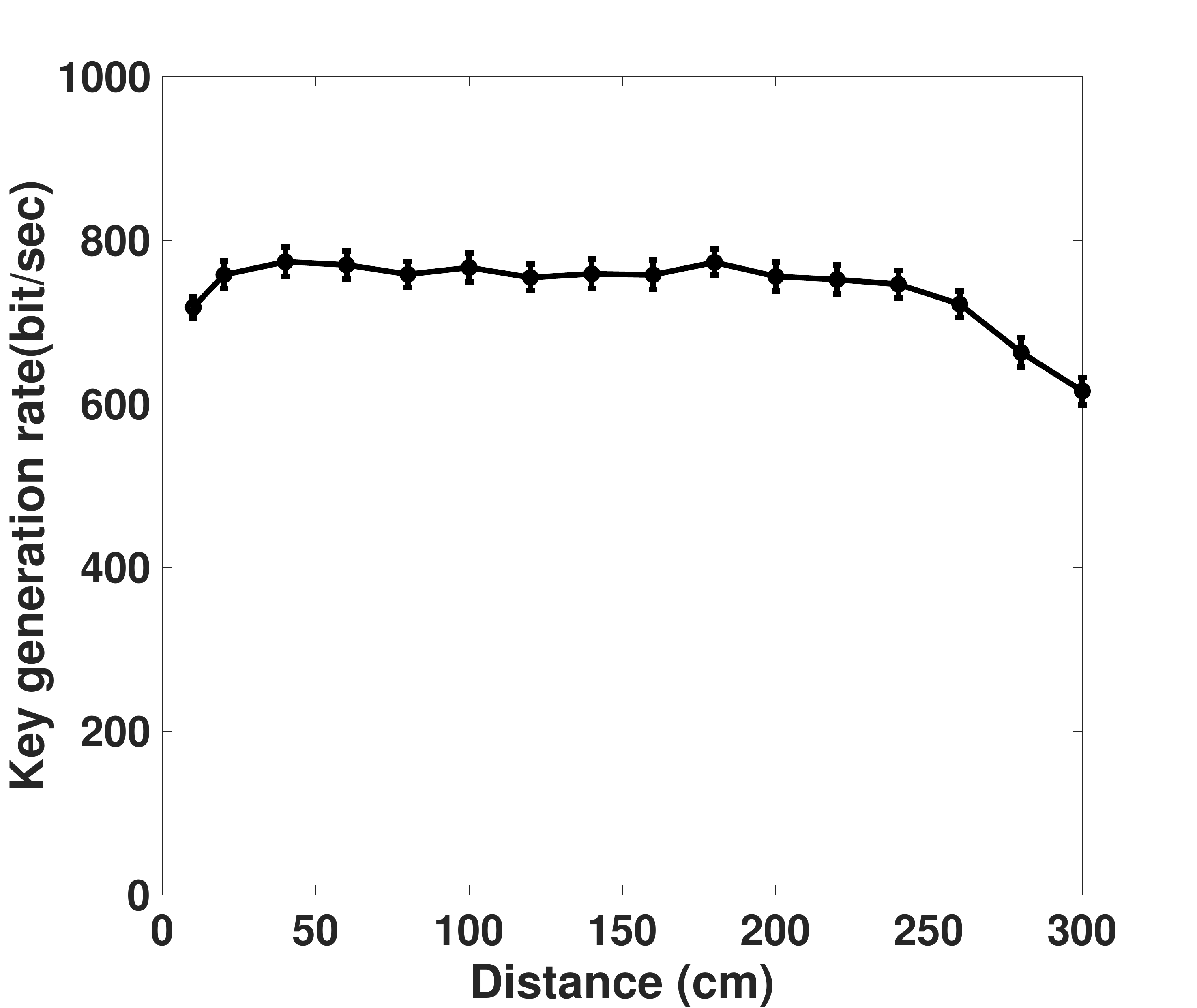}
		\label{fig:key_generation_rate_distance}}
		\hspace{-0.1in}
			\vspace{-0.1in}
	\caption{Impact of $\alpha$, $S$ and distance.}
	\label{fig:impact_of_alpha}
	\vspace{-0.2in}
	\end{figure*}
\textbf{Experimental Setup.}
We implement \SystemName on Samsung S10 which is equipped with microphone and speaker. The frequency range of the inaudible acoustic signal is 18-22~kHz, and the sampling rate is 48~kHz. The volume of the speaker is set to the maximum, and the corresponding sound pressure level (SPL) of the acoustic signal is 82~dB. Bluetooth Low Energy broadcast is used as the public channel to exchange reconciliation information. These settings can be supported by most modern mobile devices. We conduct extensive experiments with four smartphones, namely, Alice, Bob, Eve, and David (Eve's partner). Same as the state-of-the-art~\cite{lu2019free,xi2016instant}, data are collected in four different scenarios: \textbf{A}---indoor static, \textbf{B}---outdoor static, \textbf{C}---indoor mobile, \textbf{D}---outdoor mobile. In mobile scenarios, the users can shake Alice and Bob, or carry them and walk around. In static scenarios, Alice and Bob are stationary while some people are moving around. The indoor experiments are conducted in a student laboratory while the outdoor experiments are carried out on a campus road.  In each environment, we vary the distance between Alice and Bob from 10~cm to 300~cm to evaluate the impact of distance. Eve and David are located at least 10~cm away from the legitimate devices.

\vspace{-0.15in}
\subsection{Parameter Selection}
\label{subsec:parameterselection}
We first evaluate the impact of important parameters in \SystemName which include the guard band ratio $\alpha$ in quantization (Sec.~\ref{subsubsec:multiplequantizer}) and the number of eigenvectors $S$ in privacy amplification (Sec.~\ref{subsec:privacyamplification}).

The guard band ratio $\alpha$ trades off the key generation rate and key agreement rate,  Generally, a larger $\alpha$ means more samples
are discarded which improves bit matching rate but decreases
the bit generation rate. Fig.~\ref{fig:impact_of_alpha} shows the impact of $\alpha$ on key generation rate and key matching rate. First, we can see the matching rate increases significantly after information reconciliation. Also, we can see that the key agreement rate rises with the increase of $\alpha$ as more mismatch bits are discarded. In \SystemName, we use $\alpha =0.9$ because it achieves a high key agreement rate while generating sufficient bits. Although a larger $\alpha$ decreases key generation rate, from Fig.~\ref{fig:key_generation_rate_alpha} we can see that it can still achieve a generation rate of 768 bit/sec with $\alpha =0.9$. The key generation rate of \SystemName significantly outperforms the state-of-the-art works: it is 3$\times$ faster than FREE~\cite{lu2019free}, 7$\times$ faster than TDS~\cite{xi2016instant}, 30$\times$ faster than Walkie-Talkie~\cite{xu2016walkie}, 256 $\times$ faster than H2B~\cite{lin2019h2b}, respectively. Therefore, in terms of key generation rate, \SystemName is a better option than using radio signal (e.g., TDS~\cite{xi2016instant}), motion sensor signal (e.g., Walkie-Talkie~\cite{xu2016walkie}, Shake-n-Shack~\cite{shen2018shake}), and biometrics signal (e.g., H2H~\cite{rostami2013heart}, H2B~\cite{lin2019h2b}).

The number of selected eigen vectors $S$ trades off the key generation rate and entropy. If $S$ is larger, \SystemName can generate more keys but with lower entropy. If $S$ is smaller, \SystemName can generate fewer keys with higher entropy. Fig.~\ref{fig:Impact_S} shows the impact of $S$ on key generation rate and entropy. We can see that if $S$ is less than 80, the improvement of entropy levels off. Therefore, we choose the largest 80 eigenvectors to form the decorrelation matrix.
\vspace{-0.1in}
\subsection{Impact of Distance}
\label{subsec:impact_distance}
After determining $\alpha$, we evaluate the impact of distance between Alice and Bob on system performance.  Fig.~\ref{fig:key_agreement_rate_distance} shows that the key agreement rate decreases slightly when the distance between Alice and Bob increases from 10~cm to 220~cm. Then, it starts to drop quickly when the distance further increases from 220~cm to 300~cm. This is because when the distance increases, the audio signal attenuates exponentially due to path loss~\cite{soundloss}.
From Fig.~\ref{fig:key_generation_rate_distance}, we find that the key generation rate first increases when the distance increases from 10~cm to 20~cm, then becomes stable from 20~cm to 230~cm after which they start to drop rapidly. This is because Line-of-Sight channel dominates the signal when two devices are very close and hence there is not much randomness to use. However, when the communication distance is too large, more environmental noise is involved in the received signal and the signal-to-noise ratio (SNR) becomes low, which leads to more discrepancies. From Fig.~\ref{fig:key_agreement_rate_distance} and Fig.~\ref{fig:key_generation_rate_distance}, we find that [20~cm, 220~cm] is a reasonable pairing range to achieve both high agreement rate and bit generation rate. \SystemName extends the pairing distance by 3.2 times compared to FREE~\cite{lu2019free}, and 44 times compared to TDS~\cite{xi2016instant}. 

Researchers have revealed that people's social distance varies from 1.2~m to 3.7~m~\cite{hall1968proxemics}. Therefore, \SystemName can meet the pairing requirement of mobile devices in most cases. If the distance of two users is larger than pairing range (say >3m), user can either walk a few steps closer to the target or use a shorter key for temporary association, depending on the application requirement and user's preference. 
\begin{figure*}[!ht]
	\centering
	\subfigure[Impact of different environments]{
		\includegraphics[width=2.1in]{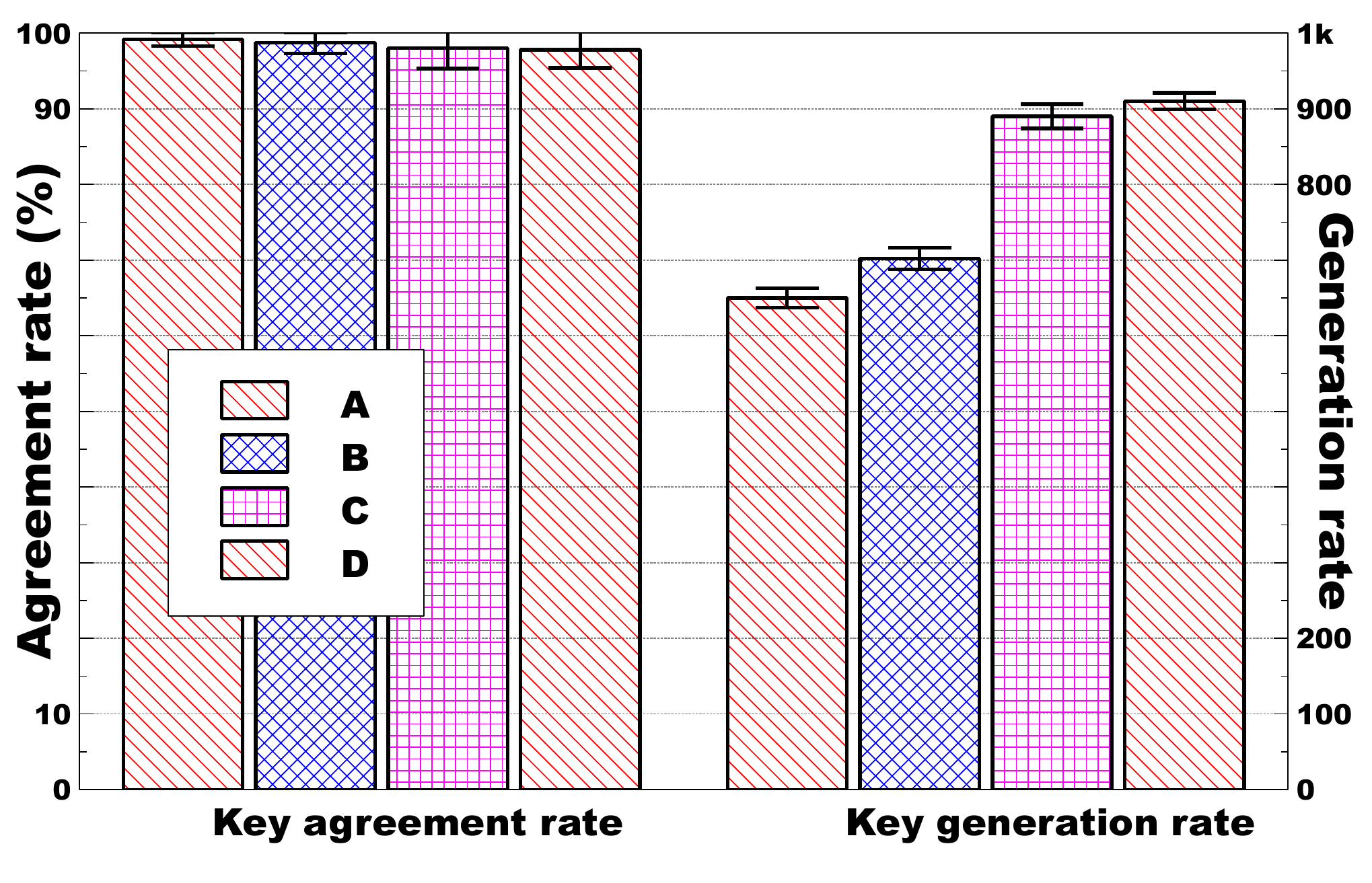}
		\label{fig:Impact_environment}}
		\subfigure[Impact of background noise]{
		\includegraphics[width=2.1in]{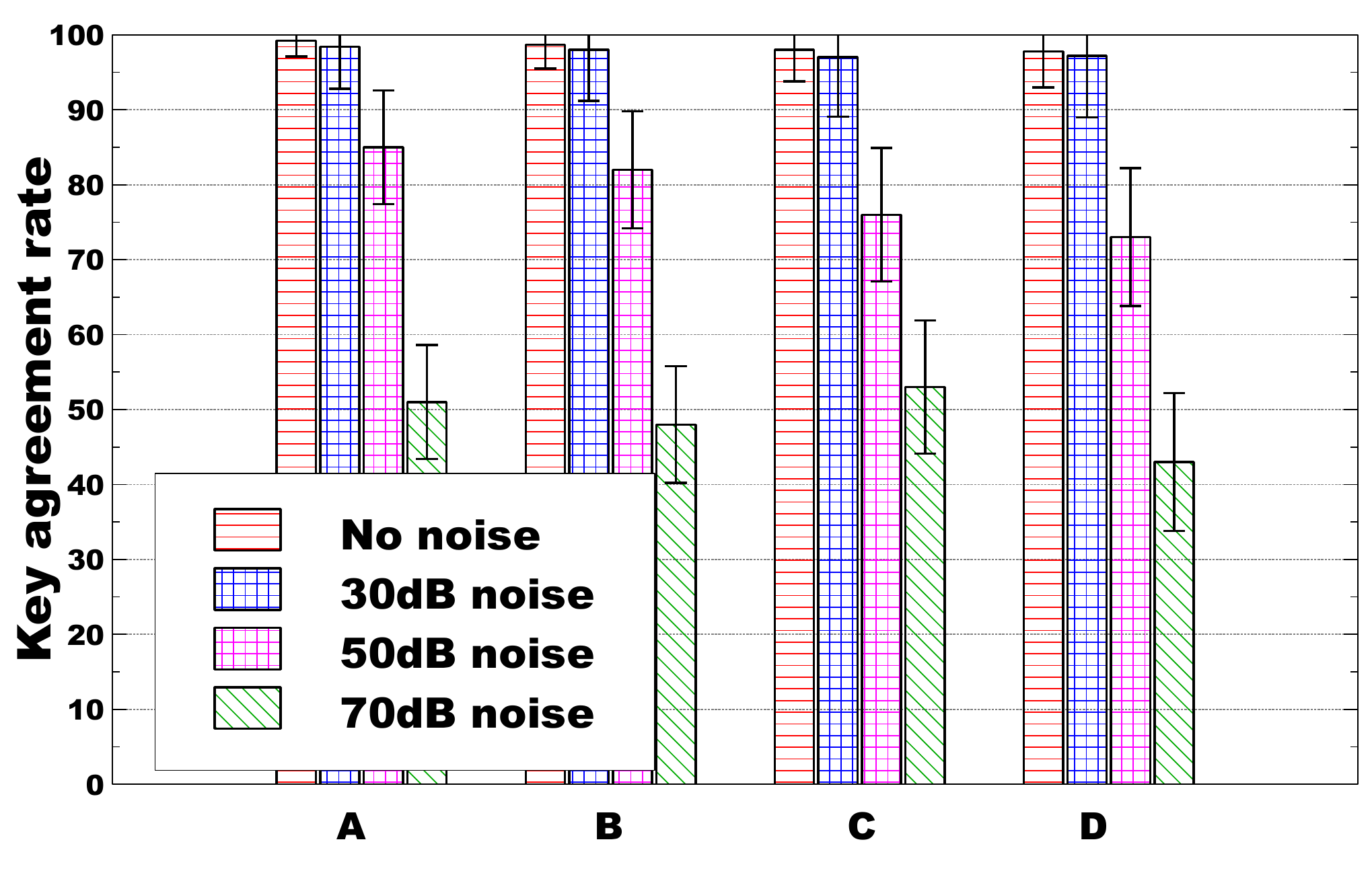}
		\label{fig:Impact_noise}}
	\subfigure[Performance in real environments]{
		\includegraphics[width=2.1in]{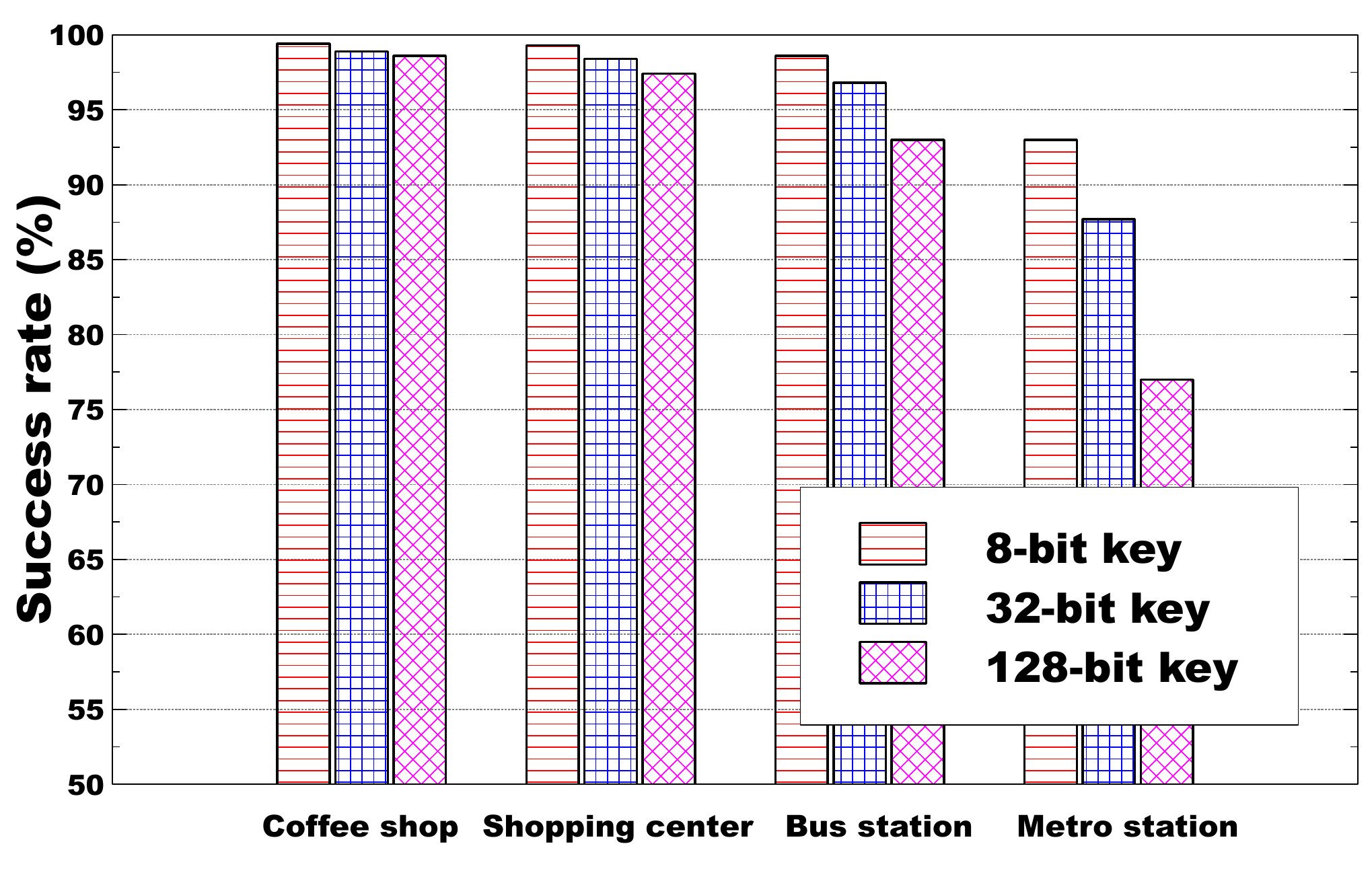}
		\label{fig:realEngironment}}
		\vspace{-0.1in}
	\caption{Impact of environment and noise.}
	\label{fig:impact_environment_noise}
	\vspace{-0.2in}
\end{figure*}
\subsection{Impact of Different Environments}
In this experiment, we evaluate the impact of different environments by using the data in the range of [20~cm, 220~cm]. Fig.~\ref{fig:Impact_environment} illustrates the key generation rate and key agreement rate in different scenarios. Intuitively, the key agreement rate of outdoor environment (i.e, B and D) is slightly lower than that of indoor environment (i.e., A and C). This is because
there are less multi-path effect and more environmental noise in outdoor environment~\cite{liu2013fast}. In terms of generation rate, we can see that the mobile scenarios (i.e., C and D) can generate more bits in comparison with static scenarios (i.e., A and B). This is because the mobile scenarios can generate more channel diversity and randomness.
\begin{figure*}[htb]
	\minipage[t]{2.2in}
	\centering
		\includegraphics[width=2.2in,height=1.2in]{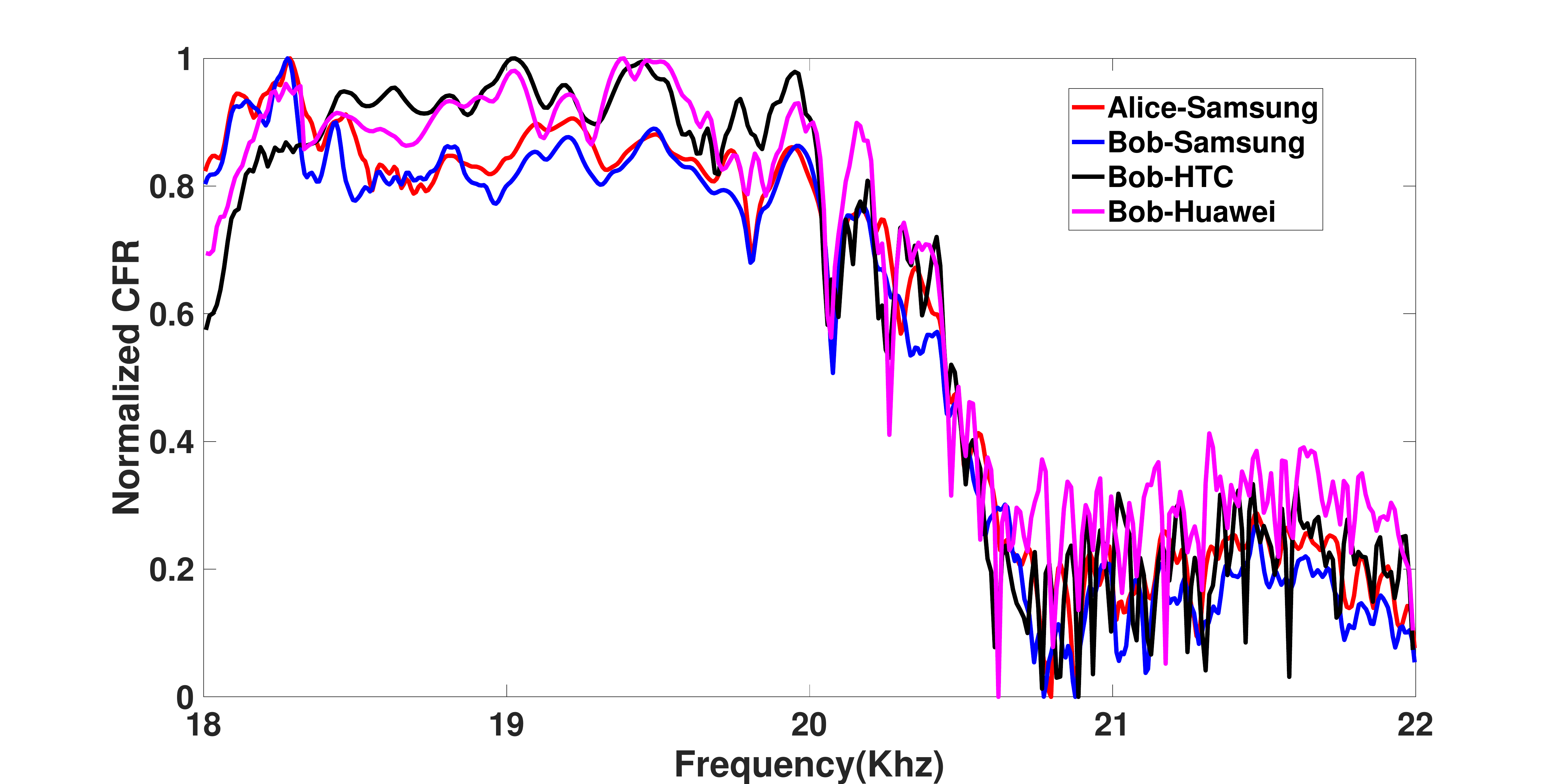}
		\captionof{figure}{CFR of different devices}
		\label{fig:different_phone}
		\endminipage
		\hfill
		\minipage[t]{2.2in}
			\centering
		\includegraphics[width=1.9in]{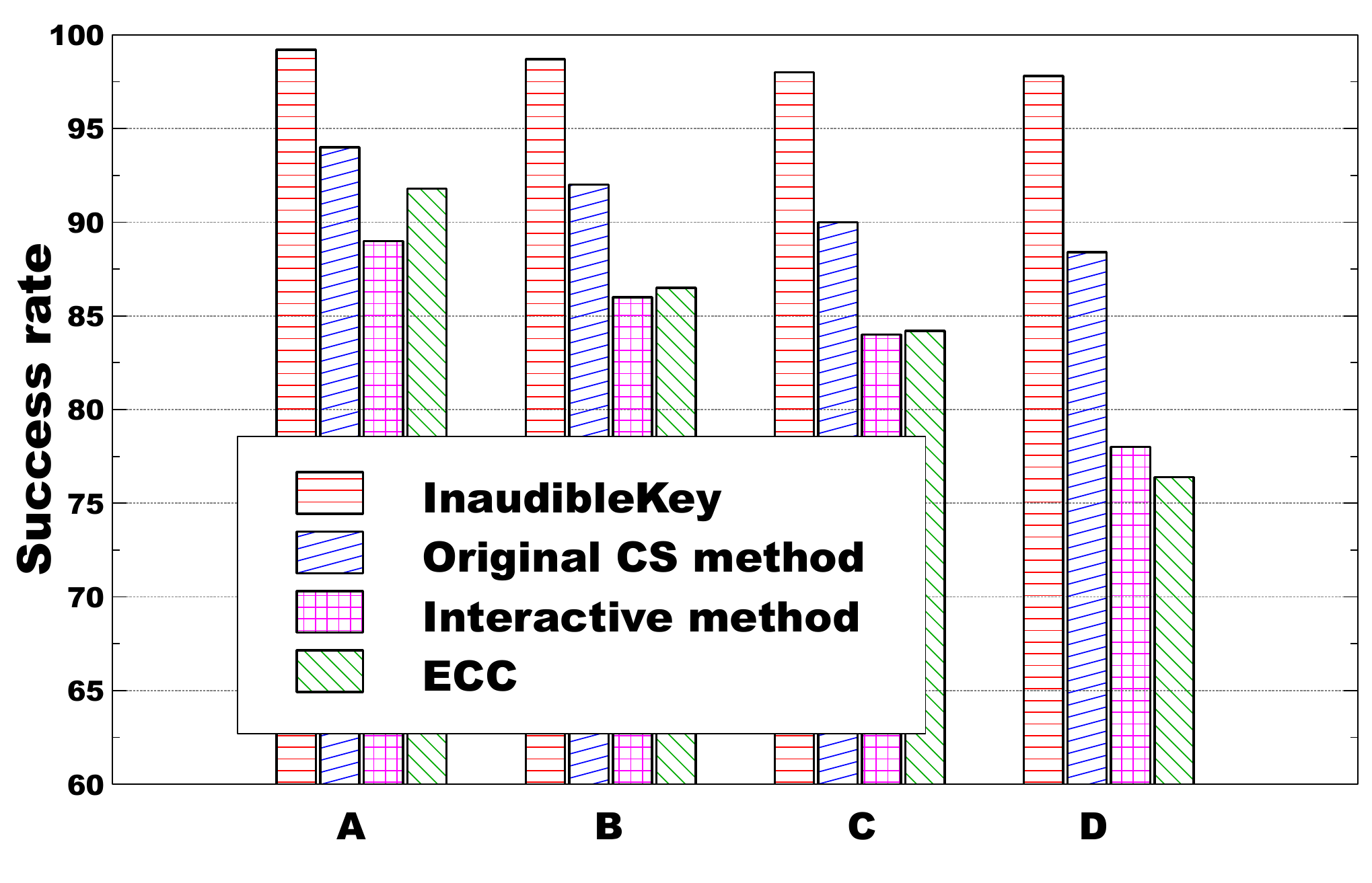}
		\captionof{figure}{\small{Different reconciliation methods.}}
		\label{fig:comparison_reconciliation}
	\endminipage
	\hfill
	\vspace{-0.2in}
	\begin{minipage}[b]{2.4in}
	\centering
\resizebox{2.4in}{!}{
\begin{tabular}{|l|l|l|l|l|}
 \hline
         & Samsung       & HTC           & Huawei        & Arduino       \\ \hline
 Samsung & \textbf{99.2\%} & 94.5\%          & 94.1\%          & 89.3\%          \\ \hline
 HTC     & 94.5\%          & \textbf{98.7\%} & 92.4\%          & 87.7\%          \\ \hline
 Huawei  & 94.1\%          & 92.4\%          & \textbf{98.8\%} & 86.4\%          \\ \hline
 Arduino & 89.3\%          & 87.7\%          & 86.4\%          & \textbf{95.6\%} \\ \hline
 \end{tabular}
 }
 \captionof{table}{Success rate of different pairs.}
\label{tab:different_device}
\end{minipage}
\end{figure*}
\vspace{-0.1in}
\subsection{Impact of Background Noise}
While our analysis above show that \SystemName consistently achieve high agreement rate, the experiments are conducted on campus only. The real-world environment is more complex and may contain various kinds of noise. We now study the degradation in matching rate with increasing background noise. We manually add 18-22~kHz random Gaussian noise with different intensities (30, 50 and 70~dB). From Fig.~\ref{fig:Impact_noise}, we can see that the key agreement rate only drops slightly when the noise level is 30~dB. However, when we further increase the noise level, the key agreement rate drops significantly. Actually, in the range of 18-22~kHz, there are little environment noise in the normal office and street environments, and such noise usually happens in factory and metro station~\cite{zhou2014acoustic}. 

To verify this, we conducted another experiment in four common environments: coffee shop, shopping centre, bus station and metro station. The distance between Alice and Bob is about 1~m, and we collect 30 minutes data for each environment. All measurements are made between 9 AM and 6 PM. In this experiment, we use success rate (the probability of generating the same key) instead of matching rate as metric because we want to know how many trials the users need to successfully pair two devices in real environments.  Fig.~\ref{fig:realEngironment} shows the success rate of each environment. We can see that \SystemName can achieve over 95\% success rate in coffee shop and shopping centre. The success rate in bus station drops slightly, but it still can achieve over 90\% success rate. We notice that the success rate in metro station drops remarkably. This is because the noise level of subway stations can be up to over 100~dB according to prior studies~\cite{lee2017analysis,gershon2006pilot}. More importantly, it has more noise in the inaudible frequency range. However, if we use 8-bit key, the success rate is still above $92\%$. The results suggest that in noisy environment, users can use a short key to improve success rate.
\subsection{Impact of Device Diversity}
So far, we assume both Alice and Bob use the same model of smartphones (i.e., Samsung S10). Now we evaluate the performance of \SystemName when Alice and Bob are different types of devices. Fig.~\ref{fig:different_phone} plots the CFRs of different types of mobile devices, we can see that when Alice and Bob are using Samsung S10, their CFRs are very close to each other. While when Bob is using HTC smartphone and HUAWEI watch,  it involves more differences, but the CFR pattern over the frequency band is still very similar. Tab.~\ref{tab:different_device} shows the key agreement rate using different devices. It is intuitive that \SystemName achieves the highest success rate when Alice and Bob are the same type of devices. The success rate drops by 4.7--9.2\% when Alice and Bob are different types of devices. This is because different types of microphone and speaker produce different impact in the transmitted and recorded acoustic signals. In particularly, the matching rate of Arduino with other devices are the lowest because we use a low-price microphone and speaker module as will be discussed in Sec.~\ref{subsec:implementation}.
\subsection{Comparison of Reconciliation Methods}
\label{subsec:comparison_reconciliation}
To demonstrate the advantage of our optimisation algorithm, we compare it with the original CS-based reconciliation method and other methods. In the literature, two commonly used reconciliation techniques are error-correcting code (ECC)~\cite{liu2012collaborative,xu2017gait} and interactive method~\cite{mathur2008radio,zeng2010exploiting,xu2016walkie,meng2014multiple}. In this paper, we use Reed-Solomon code RS(15,7) and the method in~\cite{zeng2010exploiting} as benchmark. We calculate the key agreement rate of each method and plot their results in Fig.~\ref{fig:comparison_reconciliation}. The result of original CS-based method is obtained by averaging the results of randomly generating sampling matrix 30 times. As can be seen from Fig.~\ref{fig:comparison_reconciliation}, \SystemName outperforms the original CS-based method, interactive method and ECC and  consistently achieves the highest agreement rate in all the environments. 
\begin{figure*}[!t]
	\centering
	\subfigure[Key agreement rate]{
		\includegraphics[width=1.5in]{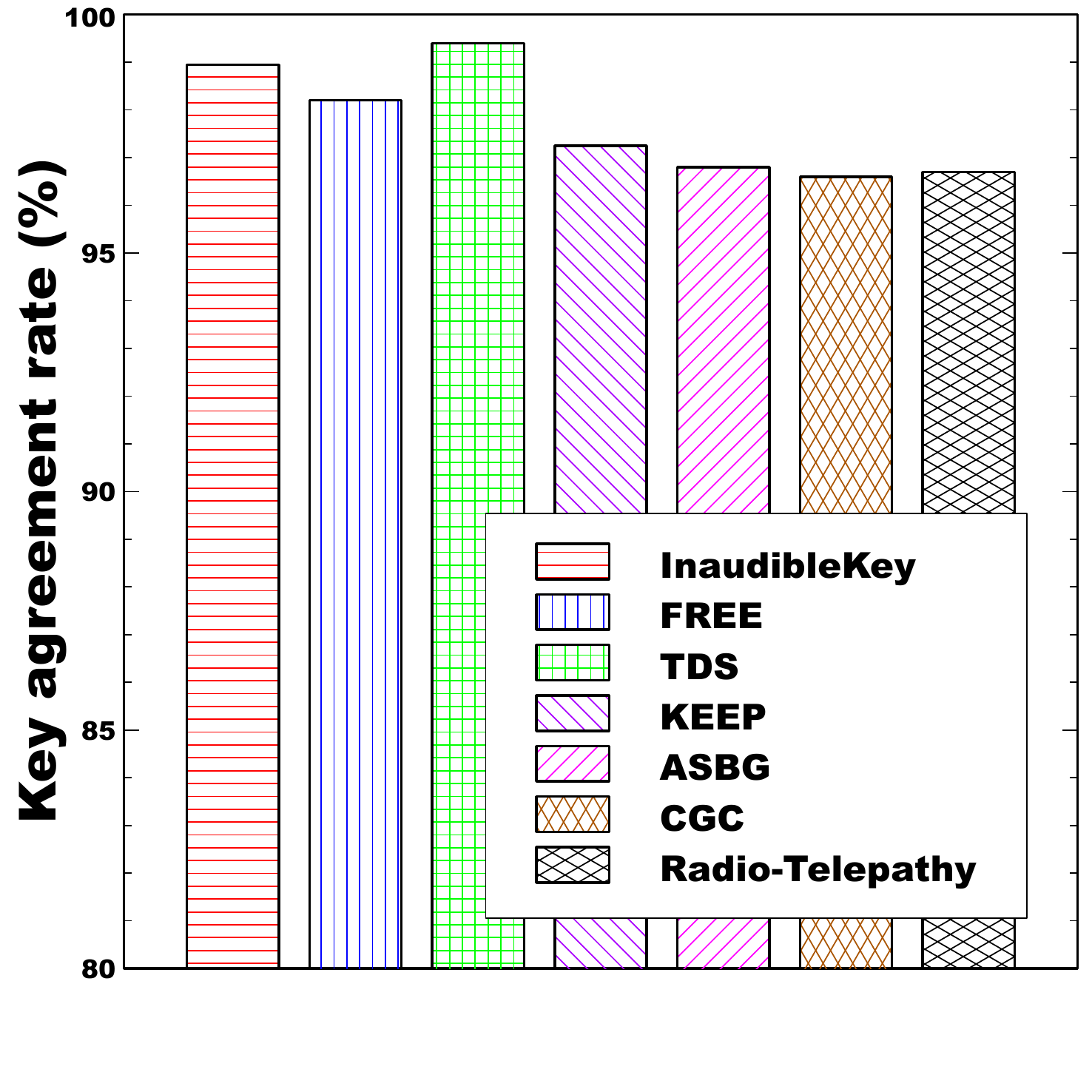}
		\label{fig:keyagreementrate_comparison}}
		\subfigure[Key generation rate]{
		\includegraphics[width=1.5in]{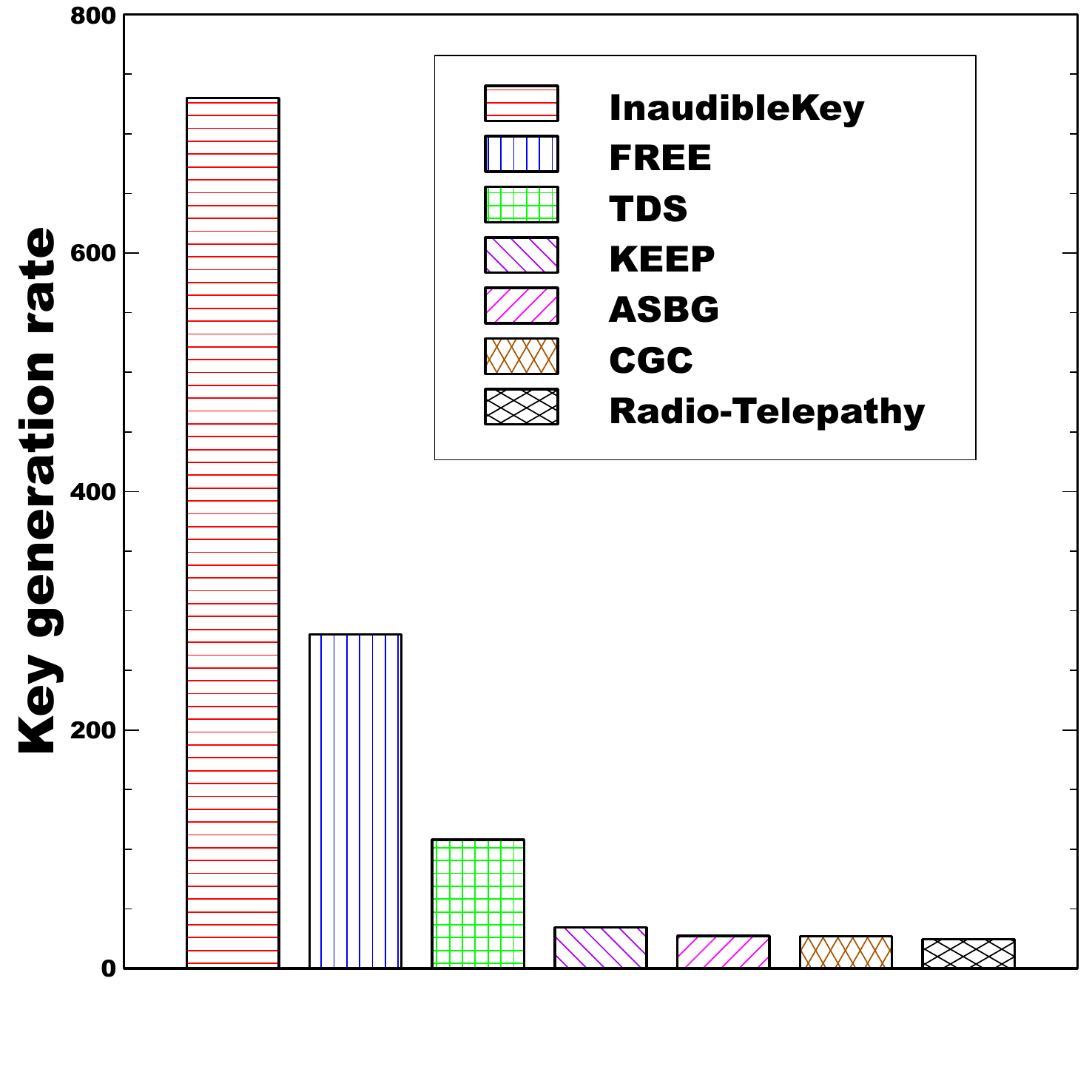}
		\label{fig:keygenerationrate_comparison}}
		\subfigure[Entropy]{
		\includegraphics[width=1.5in]{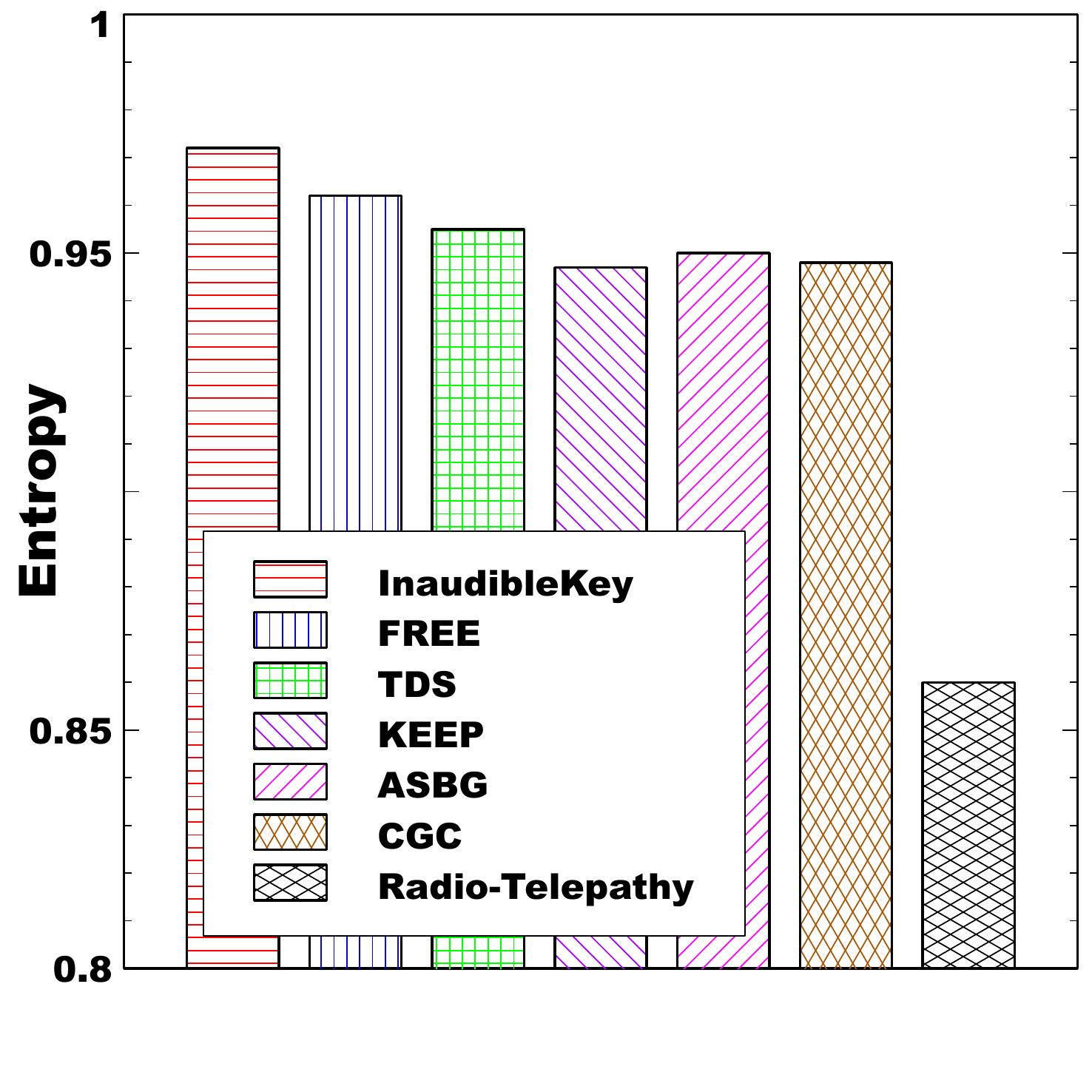}
		\label{fig:entropy_comparison}}
		\subfigure[Reconciliation counts]{
		\includegraphics[width=1.5in]{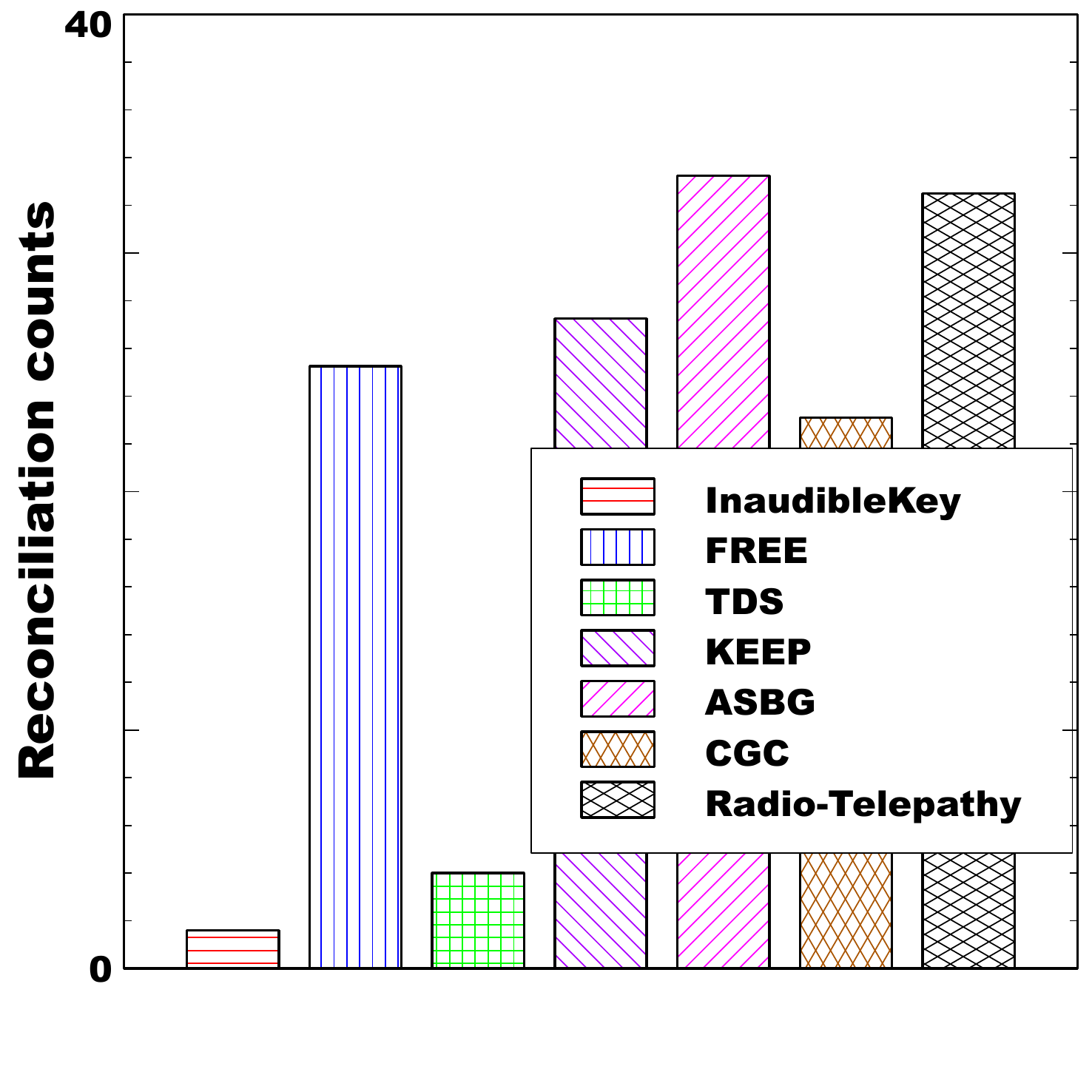}
		\label{fig:reconciliationcounts_comparison}}
		\vspace{-0.1in}
	\caption{Comparison with state-of-the-arts.}
	\label{fig:comparison}
	\vspace{-0.1in}
\end{figure*}
\vspace{-0.15in}
\subsection{Comparison with state-of-the-arts}
In this subsection, we compare \SystemName with several representative key agreement approaches for mobile networks. These methods include KEEP~\cite{xi2014keep}, ASBG~\cite{jana2009effectiveness}, TDS~\cite{xi2016instant}, Radio-telepathy~\cite{mathur2008radio},  CGC~\cite{liu2013fast} and FREE~\cite{lu2019free}. To achieve a fair comparison, we fine-tune the parameters of other methods to make sure they achieve the best performance. Specifically, for FREE, the distance between Alice and Bob is set to 80cm, and the block size is 30. For ASBG, KEEP and CGC, $\alpha$ and fragment size are set as 0.35 and 50, respectively. For TDS, the block size $\beta$ is 5. The distance between Alice and Bob is set to 4~cm as suggested by their authors. We compare the key agreement rate, key generation rate, entropy, and reconciliation counts of different methods.

Fig.~\ref{fig:comparison} shows the performances of different approaches. From Fig.~\ref{fig:keyagreementrate_comparison}, we can see that the key agreement rate of \SystemName is higher than other approaches except for TDS. In this experiment, TDS can achieve the highest agreement rate because of the short distance (4~cm). However, such short distance is unrealistic in practice. From Fig.~\ref{fig:keygenerationrate_comparison}, we can see that the key generation rate of \SystemName is significantly higher than previous works. To be specific, the key generation rate of \SystemName is 3$\times$ faster than FREE~\cite{lu2019free}, 7$\times$ faster than TDS~\cite{xi2016instant} on average. There are a number of reasons for the improvement. First, the sampling rate of the audio signal (i.e., 48kHz) is significantly higher than the radio channel probing. Second, the channel frequency response can provide more channel information compared to channel tap used in FREE~\cite{lu2019free}. Finally, the optimised CS-based reconciliation methods can recover more mismatches as demonstrated in the last subsection.

Fig.~\ref{fig:entropy_comparison} shows the entropy of extracted keys.  We find that by using KLT to decorrelate the bit sequences, \SystemName achieves higher entropy than other methods. Fig.~\ref{fig:reconciliationcounts_comparison} plots the information reconciliation counts of different methods. We can see that \SystemName requires the minimum information reconciliation counts. To successfully generate the same key, \SystemName only requires Alice and Bob to exchange reconciliation messages 1.6 times on average. In comparison, TDS requires 4 pass checks~\cite{xi2016instant}, and FREE needs to exchange 25 reconciliation information messages on average~\cite{lu2019free}. In other words, \SystemName reduces information reconciliation counts by 2.5-16 times.

The results show that \SystemName improves the key generation rate, the entropy, and reduces reconciliation counts significantly compared to the state-of-the-arts.
\vspace{-0.15in}
\subsection{Randomness of Key}
To evaluate the randomness of the extracted keys, we apply the commonly used NIST suite of statistical tests~\cite{rukhin2001statistical}. The result of NIST statistical test are p-values of different test processes which indicate whether the key is random or not. If p-value is greater than $1\%$, then the key is considered to be random. From the results in Tab.~\ref{tab:pvalue}, we find that the p-values are all larger than $1\%$, which suggests the extracted keys have good quality in randomness.
\vspace{-0.1in}
\begin{table}[!ht]
\centering
\footnotesize
\caption{Results of NIST test.}
\label{tab:pvalue}
\begin{tabular}{llllc}
\toprule
NIST TEST                & p-value     \\ \hline
Serial                   &  0.553 \\ 
FFT Test                 & 0.179 \\ 
Longest Run              & 0.353 \\ 
Monobit Frequency        & 0.742 \\ 
Linear Complexity        & 0.705 \\ 
Block Frequency          &  0.178\\ 
Cumulative Sums          &  0.741\\ 
Approximate Entropy      &  0.885\\ 
Non Overlapping Template &  0.532\\ \bottomrule
\end{tabular}
\vspace{-0.2in}
\end{table}
\begin{table}[!th]
\centering
\caption{System overhead.}
\label{tab:systemoverhead}
\resizebox{\linewidth}{0.35in}{
\begin{tabular}{cccc|ccc}
\hline
                                                                  & \multicolumn{3}{c}{Samsung S10} & \multicolumn{3}{c}{Arduino}    \\ \cline{2-7} 
                                                                  & InaudibleKey & RSA & ECDHE-RSA & InaudibleKey & RSA  & ECDHE-RSA \\ \hline
\begin{tabular}[c]{@{}c@{}}Processing\\ Time (ms)\end{tabular}    & 124          & 361 & 347       & 891          & 4,196 & 5,481      \\ \hline
\begin{tabular}[c]{@{}c@{}}Energy\\ Consumption (mJ)\end{tabular} & 108          & 391 & 354       & 1,107         & 1,706 & 2,196      \\ \hline
\end{tabular}
}
\end{table}
\vspace{-0.1in}
\begin{figure}[!th]
	\centering
\includegraphics[width=2.2in]{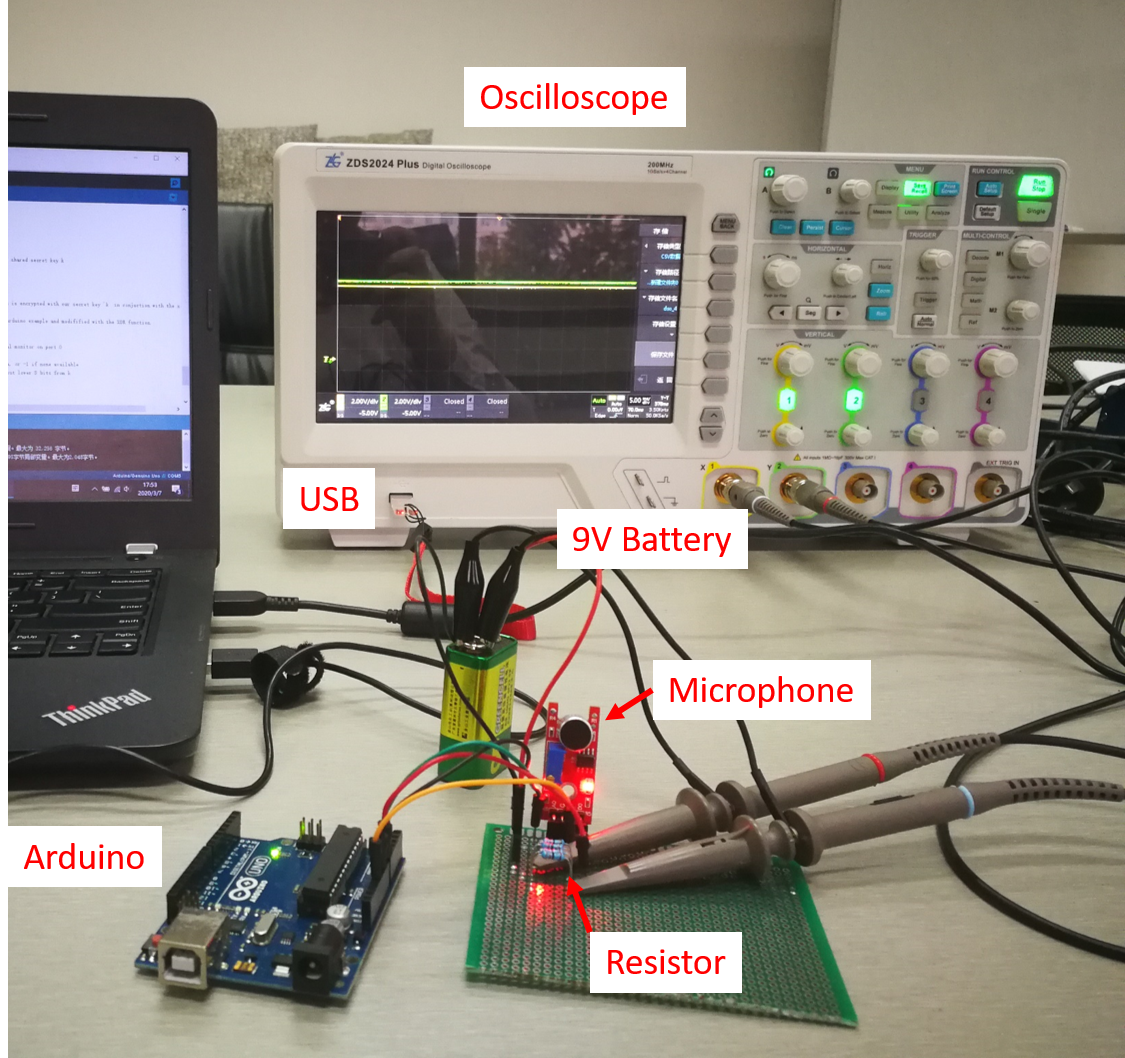}
	\caption{Experimental setup of energy consumption.}
	\label{fig:poweranalysis}
\end{figure}

\subsection{System Implementation}
\label{subsec:implementation}
To validate the feasibility of \SystemName on various IoT devices, we implement the prototype of \SystemName on Samsung S10 smartphone and Arduino Uno board. 
\begin{figure*}[!th]
	\centering
		\subfigure[Agreement rate of different attacks]{
		\includegraphics[width=1.8in]{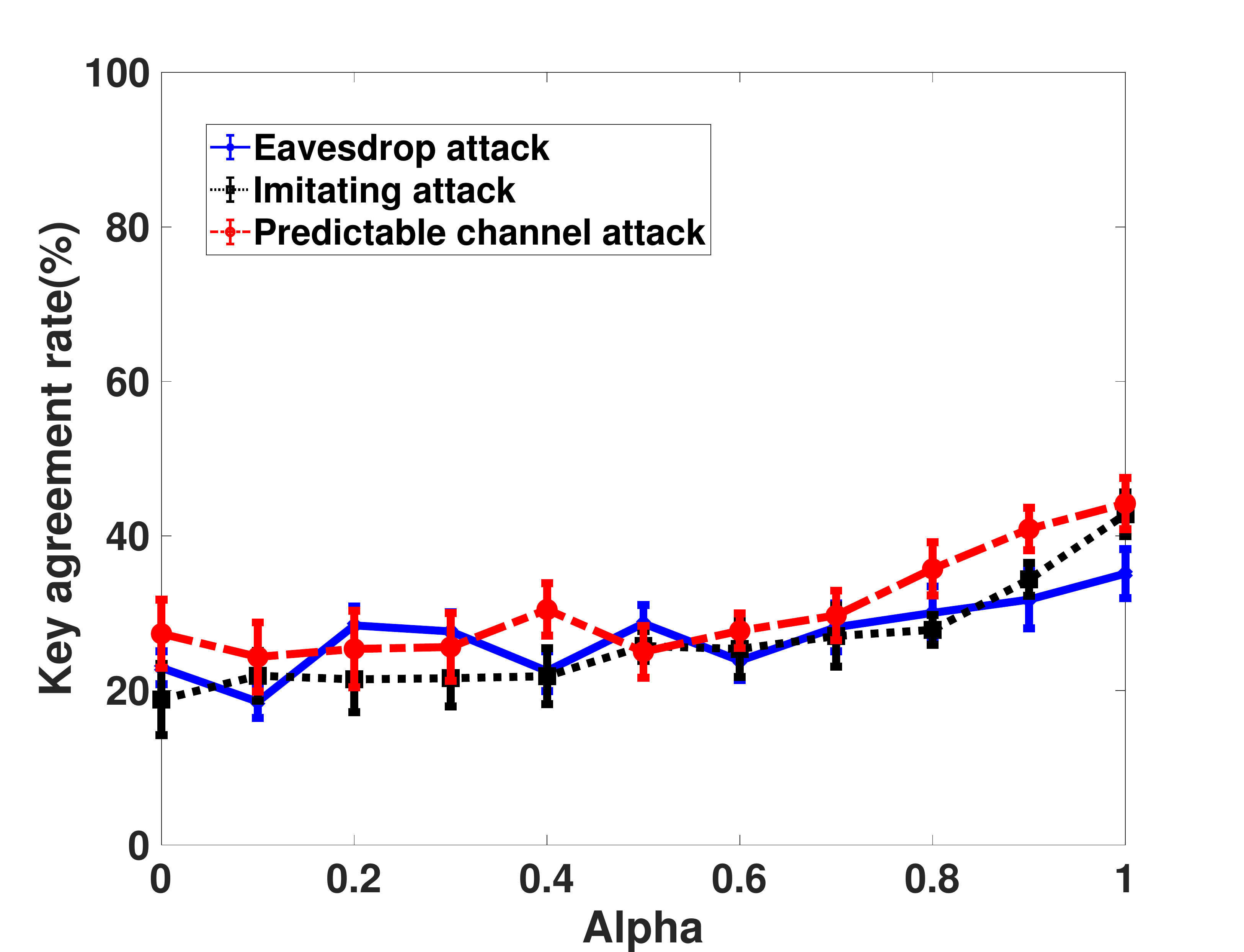}
		\label{fig:Agreement_attacker}}
			\subfigure[Bob vs predictable channel attacker]{
		\includegraphics[width=1.8in]{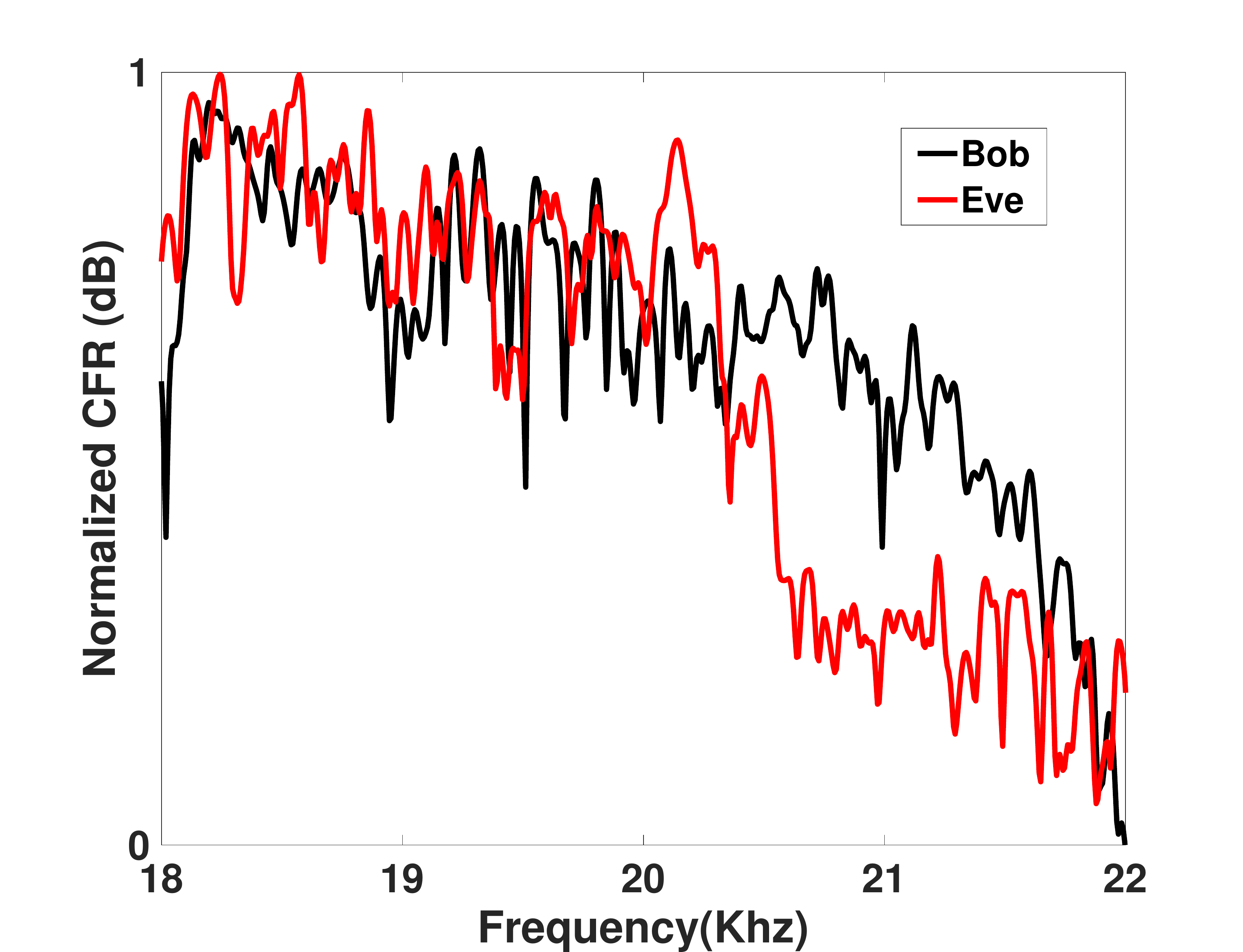}
		\label{fig:PredictableAttack_CFR}}
		\subfigure[Feasible range of parameter M]{
		\includegraphics[width=1.8in]{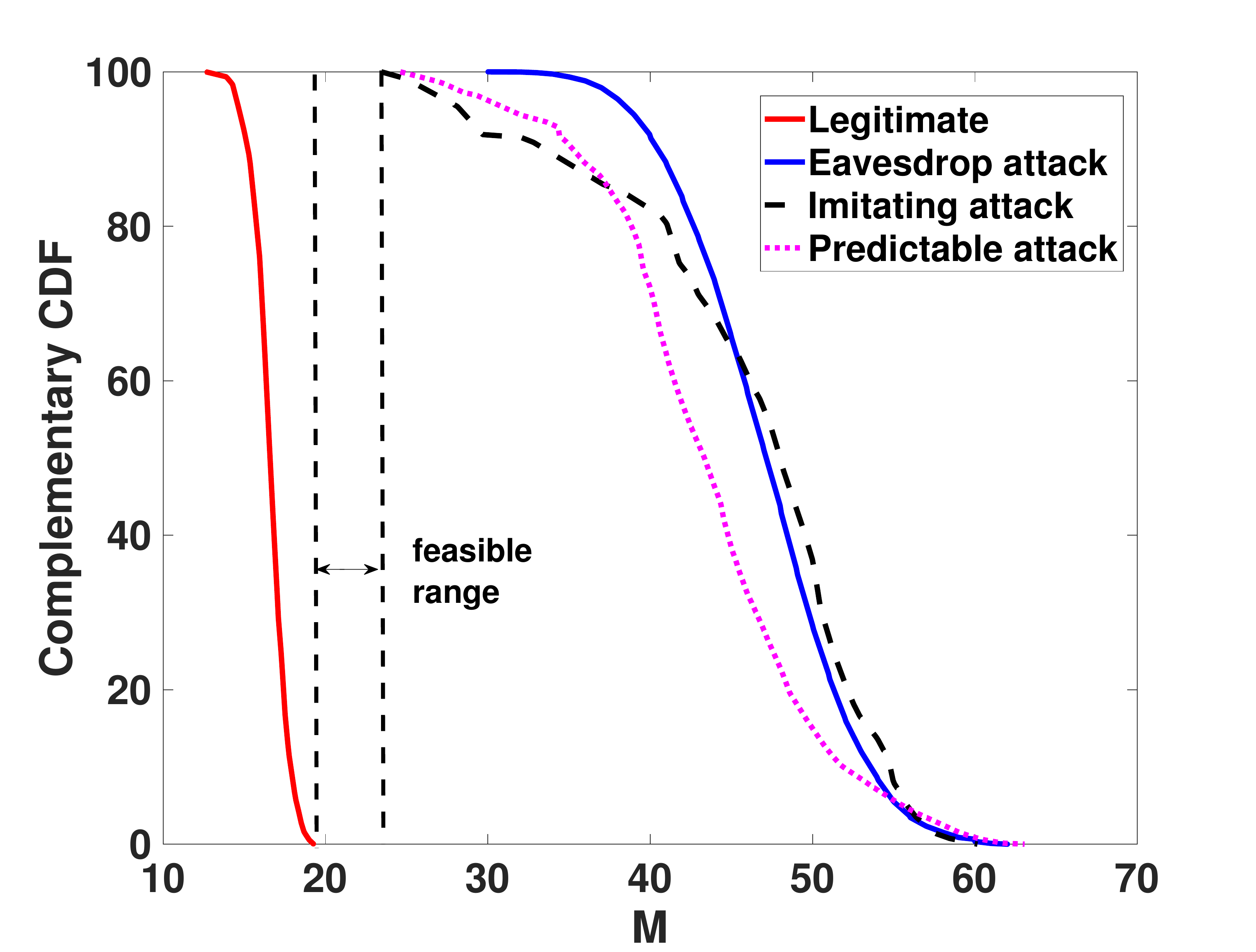}
		\label{fig:SecrecyofCS}}
		\vspace{-0.15in}
	\caption{Security analysis.}
	\label{fig:securityanalysis}
	\vspace{-0.2in}
\end{figure*}

The CPU of Samsung S10 is an Snapdragon at 2.84~GHz and the operating system is Android 9.0. It is equipped with a stereo speaker and two dedicated microphone with active noise cancellation function. Only the bottom microphone is used because it is close to the speaker. The system is implemented in Java and the MAC algorithm described in Section~\ref{sub:reconciliation} is implemented based on SHA256 (HMAC-SHA256). In \SystemName, we save the transmitted OFDM signal as a Waveform Audio (WAV) file with a format of 16-bit Pulse Coded Modulation (PCM), which will be played by speaker. To reduce the expected response time, we implement \SystemName in multiple threads. Two threads are created after \SystemName is launched. One of the threads is responsible for transmitting WAV file. After transmitting, the smartphone will transit into listening mode and another thread which records audio signal from another phone will be created. In reconciliation, \SystemName uses \emph{$\ell_1$-Homotopy}~\cite{DonohoTsaig:08}  which is an efficient implementation of $\ell_1$ optimization algorithm.  The complexity of \emph{$\ell_1$-Homotopy} is $O(k^3+kmn)$, where $k$ is the sparsity of the solution, $m$ and $n$ are the size of sampling matrix $A$ which is 23 and 128, respectively.

Firstly, we compare \SystemName with public key cryptography and Diffie-Hellman key exchange protocol. For public key cryptography, we use the commonly used RSA as benchmark. Alice can use RSA to encrypt a 128 bits key which can be decrypted by Bob. Then, Alice and Bob can use AES-128 to secure their communication. In this experiment, we use 2048 bits key for RSA which is recommended by NIST~\cite{NISTKey}. Traditional Diffie-Hellman protocol is susceptible to MITM attack, and is rarely used in practice. Therefore, we use the commonly used Elliptic Curve Diffie Hellman Ephemeral with RSA signature (ECDHE$\_$RSA), which is used in Transport Layer Security (TLS). We implement these algorithms on Samsung S10 and calculate their processing time and energy consumption. The implementation of these cryptographic algorithms are based on the Chilkat library~\footnote{http://www.chilkatsoft.com/}. The computation time is obtained from the console of the development environment (Android studio) and averaged by the results from 30 tests. The energy consumption of smartphone is calculated by reading the voltage and current level of the battery which can be obtained by Android API~\footnote{https://developer.android.com/reference/android/os/BatteryManager.html}. As the results in Table~\ref{tab:systemoverhead}, we can see that RSA requires 361~ms to finish a round of encryption and decryption with a 2048 bits key. It takes about 347~ms for ECDHE$\_$RSA to generate a 128 bits key. However, \SystemName only requires 124~ms to generate a 128 bits key. Therefore, InaudibleKey is superior to public key cryptography and D-H protocol in key distribution on mobile devices.

Secondly, to verify the feasibility of \SystemName on resource-limited IoT devices, we implement our system on Arduino Uno board. Compared to the powerful CPU in Samsung S10, the microcontroller of Arduino is ATmega328 which only has 32~K flash memory and 2~K SRAM. There is no default speaker and microphone on Arduino board, so we connect additional speaker module and microphone module to it. To measure the energy consumption on Arduino, we connect the output of a 9V battery~\footnote{The operating voltage of Atmega328p is 5V, but the input voltage of the Arduino board is 6V to 12V.} to digital oscilloscope. The details of the experimental setup is shown in Fig.~\ref{fig:poweranalysis}. The voltage over the resistor is stored in USB and used to calculate the energy consumption of the board. The processing time and energy consumption is shown in Table~\ref{tab:systemoverhead}. We can see that  although the system overhead of \SystemName is much higher than that on smartphone, it is still much more efficient than RSA and ECDHE-RSA. Moreover, the computation-intensive part of \SystemName---reconciliation--- can be performed on the power-rich device (if one of the devices is powerful).

We now analyse the impact of energy consumption on IoT devices. The battery capacity of the Samsung S10 is 3,400~mAh (42.8~kJ). So, the energy cost of \SystemName amounts to 0.3$e^{-5}$ of the total energy supply. If we assume the smartphone with a targeted lifespan of one day which results in an energy budget of 1.75~kJ per hour. Then, with only $1\%$ of the battery budget (17.5~J), \SystemName is able to run approximately 175 times per hour, i.e., \SystemName can
continuously run every 20 seconds. In the same way, we can estimate that with 9V battery (500~mAh) and $1\%$ of the battery budget, \SystemName is able to run every 22 minutes on Arduino Uno board, i.e., it can run about 3 times per hour. These results demonstrate that \SystemName incurs a low system overhead and is more efficient than public key scheme.
\section{Security Analysis}
\label{sec:securityanalysis}
\subsection{Against Vulnerability 1}
In Vulnerability 1, Eve can try to reconstruct the keys from $y_{Bob}$ directly using $\ell_1$ optimization. As discussed above, the key generated by \SystemName has high entropy, which means almost half of the keys are bit `1's. Fig.~\ref{fig:Impact_environment} shows that the initial agreement rate of Alice and Bob is about $84\%$ when $\alpha=0.9$. Assume we use 128-bit key, we have $P=128\times 0.84 \times log(128/(128\times0.84))\approx 19$ and $Q=128\times 50\% =64$ according to~
\cite{lin2019h2b}. Theoretically, the range of M can be [20,63] because $P<M<Q$. Practically, we can choose $M \in [23,50]$ to guarantee security against an adversary and the availability of the same key.  
\subsection{Against Vulnerability 2}
Eve can perform the following three types of attacks to generate a key $K_{Eve}$ that is close to $K_{Bob}$ hoping that she can recover $K_{Bob}$ with the eavesdropped $y_{Bob}$. 
\paragraph{Against Eavesdropping Attack}
In this attack, Eve can eavesdrop all the communication traffic in the public channel. Since Eve is located out of the safe distance (>10~cm), she will obtain a totally different channel response, as discussed in Sec.~\ref{sec:preliminarystudy}. From Fig.~\ref{fig:Agreement_attacker}, we can see that the agreement rate of eavesdropping attack is about 20-35\%. Therefore, if Eve is out of the safe distance, she cannot guess the same key due to the different multipath fading channel.
\paragraph{Against Imitating Attack}
In this attack, Eve can observe how Alice and Bob generate keys. Then, after Alice and Bob leave the site, Eve will ask her partner David to imitate the motion of Alice and Bob to generate the same key. Previous studies have shown that simply imitating the user's shaking or walking motions cannot generate the same key for accelerometer-based authentication systems~\cite{xu2016walkie,shen2018shake,mayrhofer2009shake}. Similarly, Fig.~\ref{fig:Agreement_attacker} shows that an imitating attack can achieve a higher agreement rate when $\alpha$ increases. But eventually, it can at most achieve $42\%$ agreement rate. More importantly, Eve does not know which bit is correct because of the time-varying nature of channels. 
\paragraph{Against Predictable Channel Attack}
Predictable channel attack is a simple but effective attack to compromise a key agreement protocol, especially for RSSI-based approaches~\cite{jana2009effectiveness,liu2013fast}. In this attack, Eve can intentionally block and unblock the Line-of-Sight (LOS) between Alice and Bob to generate predictable channel measurements.  We evaluate this attack by setting up Alice and Bob 100~cm away with LOS and ask a person to walk between Alice and Bob intentionally. Then after the key generation, we replace Alice and Bob with Eve and David and ask the same person to repeat the process. Then we compare Eve's key with Bob's key to see if Eve can generate the same key. From Fig.~\ref{fig:Agreement_attacker}, we can see that a predictable channel attack can achieve the highest matching rate among these three types of attack. But still, it can at most reach $43\%$ matching rate. Fig.~\ref{fig:PredictableAttack_CFR} plots the CFR of Bob and Eve when the same person blocks the LOS signal. We can see that although the channel responses of Bob and Eve are similar in some frequencies, there is still a large portion of the difference in other frequencies due to time-varying channels and hardware difference. Particularly, we noticed that Eve is capable of producing similar channel responses in the lower frequency range but not the higher frequency range. There are two reasons. First, the microphone actually works as a low-pass filter with a 22~kHz cutoff frequency~\cite{he2019canceling}. So in the higher frequency range, the acoustic signal will be attenuated slightly which results in more mismatches. Additionally, Zhou et al.~\cite{zhou2014acoustic} found that different speakers’ performances are much more diversified at a higher frequency range. If the attacker leverages more sophisticated hardware, it is possible that they increase their attacking ability. But it is an open question that requires further investigation. 

Although imitating attack and predictable channel attack can achieve approximately $43\%$ matching rate, the probability of deducing the same 128-bit key is extremely low, i.e., $0.43^{128}=1.21e^{-47}$. The matching rate of Eve can be further reduced by setting a higher threshold in quantization or turning down the volume of speaker. Considering the matching rate of Eve, a 225-bit key of our system is equivalent to a 128-bit AES symmetric key, and it takes about 0.3~s to generate such a key based on the result in Sec.~\ref{subsec:impact_distance}.

Fig.~\ref{fig:SecrecyofCS} shows the distribution of $P$ and $Q$ in our dataset. We can see that there is a feasible range to use. In other words, if $M$ lies in the feasible range, then \SystemName is resilient to the three types of attacks above. Previous studies also found that if the same sampling matrix $A$ is used repeatedly, both $y_{Bob}$ and $K_{Bob}$ could be conditionally accessed~\cite{yang2015security}. We can easily solve this problem by updating $A$ after each successful key generation. Although $A$ needs to be pre-stored, it is public information instead of a secret that is only known by Alice and Bob.

\SystemName can be used to facilitate the pairing process of two devices but cannot authenticate their identities. Similar to many previous studies~\cite{xie2018genewave,xi2016instant}, our assumption is the devices within the pairing distance are legitimate. However, since \SystemName achieves much longer working distance than previous systems, it is possible that an attacker approaches the user and pretends to be a legitimate device to launch key generation. In this case, we have to either involve user in the loop or adopt traditional authentication approaches, such as pre-shared key or token-based methods.

\vspace{-0.1in}
\section{Related Work}
\label{sec:relatedwork}
\textbf{Proximity-based approaches.}
The proximity-based approaches pair two devices based on the observation that two devices in physical proximity can measure similar physical information. Researchers have proposed many different systems by exploring various location-sensitive features such as RSSI~\cite{varshavsky2007amigo,mathur2011proximate,zhang2017proximity}, CSI~\cite{xi2016instant}, audio~\cite{schurmann2011secure} and illumination~\cite{miettinen2014context}. 
However, these approaches suffer from a common problem: the distance between two legitimate devices should be very close, e.g., 1.25~cm in Proximate~\cite{mathur2011proximate} and 5cm in TDS~\cite{xi2016instant}.

\textbf{Channel reciprocity-based approaches.}
Physical layer key generation is a hot research filed over the past decade. Researchers have studied key agreement for different wireless technologies such as ZigBee~\cite{jana2009effectiveness}, Wi-Fi~\cite{mathur2008radio,xi2016instant}. Among these approaches, RSSI-based key generation methods suffer from predictable channel attack and low bit generation rate. Although CSI-based key generation methods can improve bit generation rate, most systems rely on customised hardware to obtain CSI information.  Recently, researchers also use unique body channel to pair two mobile devices~\cite{roeschlin2018device,yang2016secret}.  
However, these methods require specialised sensors such as electrode~\cite{roeschlin2018device}  and Electromyogram sensor~\cite{yang2016secret}.


\textbf{Acoustic signal-based approaches.} 
Recently, the acoustic signal is also exploited to pair mobile devices~\cite{bala2020phy,lu2019free,xie2018genewave,han2018you,schurmann2011secure}. Proximity-based schemes such as~\cite{schurmann2011secure,xie2018genewave} are not feasible due to constraint of social distance. Two recent works~\cite{lu2019free,bala2020phy} are closely related to our system. FREE~\cite{lu2019free} used channel tap and the authors of~\cite{bala2020phy} used sound pressure as channel characteristics. However, these metrics can only provide a coarse estimation of acoustic channel. In comparison, we modulate the audio signal using OFDM technology to obtain fine-grained channel estimation and propose an optimisation algorithm to improve the performance of reconciliation. This is why we can achieve much higher generation rate. 

\section{Conclusion}
\label{sec:conclusion}
In this paper, we propose a novel key generation system for mobile devices via inaudible acoustic signal. 
Extensive evaluation results show that \SystemName outperforms the state-of-the-arts significantly. To demonstrate the feasibility, we implement \SystemName on both powerful and resource-limited IoT devices.  We also verify the security of \SystemName against malicious attacks. The results in this paper show that \SystemName is a fast, practical and efficient key generation protocol for mobile devices that can work in various environments. More importantly, it allows users to pair two mobile devices without breaking social distance restrictions.

\begin{acks}
This work was partially supported by the APRC grant (Project No. 9610485), and the Start-up grant (Project No. 7200642) from City University of Hong Kong, the project JCYJ20190808183203749 supported by the Science Technology and Innovation Committee of Shenzhen Municipality, the GRF grant from Hong Kong RGC (CityU 11217817, 9048178), and CityU SRG-Fd 7005050. The work described in this paper was partially supported by a grant from Chow Sang Sang Group Research Fund sponsored by Chow Sang Sang Holdings International Limited (Project No. 9229062). Albert Y. Zomaya would like to acknowledge the support of the Australian Research Council Discovery scheme (grant DP200103494).
\end{acks}

\small{
\bibliographystyle{unsrt}
\bibliography{main}
}

\end{document}